\documentclass[a4paper,11pt]{article}
\usepackage{jheppub}
\usepackage{lineno}
\usepackage{amsmath}
\usepackage{amssymb}
\usepackage{graphicx}
\usepackage{multicol}
\usepackage{subfigure}
\usepackage{epsfig,color}
\usepackage{braket}
\usepackage{float}
\usepackage{array}
\usepackage{caption}
\usepackage{setspace}
\usepackage{tikz-feynman}
\usepackage{slashed}
\usepackage{array}
\usepackage{booktabs}
\usepackage{multirow}
\usepackage{eufrak}
\nolinenumbers

\title{\boldmath Higgs couplings in SMEFT via Zh production at the HL-LHC}

\author[a]{Subhaditya Bhattacharya,}
\author[b]{Sanjoy Biswas,}
\author[a]{Abhik Sarkar}
\affiliation[a]{Department of Physics, Indian Institute of Technology Guwahati, Assam 781039, India}
\affiliation[b]{Department of Physics, Ramakrishna Mission Vivekananda Educational and Research Institute,
Belur Math, Howrah 711202, India}

\emailAdd{subhab@iitg.ac.in}
\emailAdd{sanjoy.phy@gm.rkmvu.ac.in}
\emailAdd{sarkar.abhik@iitg.ac.in}

\abstract{We study the Higgs couplings involved in the $Zh$ associated production mode at the Large Hadron Collider (LHC) in presence of Higgs-gauge boson coupling modifiers via $\kappa$ framework, and dimension 6 Standard Model Effective Theory (SMEFT) operators. The analysis is performed mainly in context of the HL-LHC (with $\sqrt{s}=$14 TeV and luminosity 3000 $fb^{-1}$) setup using cut based as well as machine learning techniques. The analysis shows significant betterment in the signal significance by using the machine learning technique. We also do a $\chi^2$ analysis, which reveals an appreciable change in the sensitivity of the coupling modifiers due to the presence of effective operators, in particular due to the $qqZh$ Higgs-current interactions. The dipole operators concerning the same vertex, combine quarks of different chiralities and does not interfere with the SM diagrams, hence contributing only at $\mathcal{O} (\Lambda^{-4})$. In order to observe better sensitivity for this class of operators, we need to move to higher center-of-mass energy, where the effects of both Higgs-current and dipole 4-point interactions are more prominent.}


\begin{document}
\maketitle
\flushbottom


\section{Introduction} \label{sec:intro}

The Higgs boson discovery at the LHC provided the first conclusive portrait of the Electroweak Symmetry Breaking (EWSB) mechanism \cite{Logan:2014jla} and mass generation of the SM particles. Over a decade since the Higgs discovery, we have been able to probe many of its properties, some of which are precise and closely mimic that of a SM-like isodoublet ($H$), some are not, allowing for New Physics (NP) studies via Higgs sector. One standard way of parametrizing the Higgs couplings beyond the SM is to use the $\kappa$ framework \cite{ParticleDataGroup:2022pth}, where any deviation from the SM is taken into consideration by variation of the $\kappa$ parameter, where $\kappa = 1$ depicts SM. For example, concerning the $hZZ$ coupling, the corresponding $\kappa_{Z}=\frac{g_{hZZ}}{g^{SM}_{hZZ}}$. Most of the Higgs studies at the LHC uses such parametrisation and so do we. However, one needs to be vigilant about the possible sources of NP that generates such contributions and need to be consistent with null observation of NP searches while scanning the possible values that $\kappa$ can acquire\footnote{We will address this issue later in a specific context.}.

Over the last two runs, LHC experiments, ATLAS \cite{ATLAS:2022vkf} and CMS \cite{CMS:2022dwd} have almost pinned down the Higgs couplings to less than 10\% uncertainty, however, the High Luminosity (HL) run will prove to be significant for establishing a clearer picture of Higgs characteristics. The Higgs couplings associated with gauge bosons has been studied extensively in the existing literature from various point of views. They include bounds on the anomalous couplings \cite{CMS:2019jdw, Anderson:2013afp, Sharma:2022epc,Hernandez-Juarez:2023dor,Dutta:2008bh}, Higgs CP properties \cite{Desai:2011yj, Christensen:2010pf, Dwivedi:2016xwm,Bernlochner:2018opw,Brehmer:2017lrt}, etc. The procedure involves dedicated signal observables as well as in the effective theory (EFT) framework \cite{Banerjee:2015bla, Banerjee:2018bio, Banerjee:2019pks, Banerjee:2019twi, Freitas:2019hbk, Bishara:2022vsc,Bishara:2020pfx,Englert:2024ufr,Banerjee:2020vtm, Banerjee:2021huv, Biswas:2021qaf, Araz:2020zyh,Li:2023tcr,Yan:2021tmw,Xie:2021xtl} in variety of production and decay channels in context of the current and future colliders.

In this analysis, we study the NP effects in $hZZ$ and $qqZ$ couplings as well as the fully NP induced $qqZh$ coupling in the $Zh$ associated production mode 
at the High Luminosity (HL) LHC frontier. We adhere to the $\kappa$ frameowrk to depict NP contribution coming to $hZZ$ coupling, whereas the NP contributions to $qqZ$ and $qqZh$ are studied within the realm of the Standard Model Effective Field Theory (SMEFT) \cite{Brivio:2017vri} . SMEFT framework is arguably the best motivated scenario for searching NP in a model independent way, either via deviations in the signal pattern or excess in the cross-section predicted from the SM, given the absence of a specific hint for NP in current scenarios. We would also like to note here, that while the contributions to $\kappa_Z\neq 1$ may very well appear from the SMEFT operators, however, there are restrictions like Z mass corrections, which may constrain such contributions. Therefore, parametrising the $hZZ$ coupling via $\kappa_Z$ makes us explore possibilities even beyond SMEFT. In short, by using both $\kappa$ framework and SMEFT, we are basically taking the leading contributions to the relevant couplings that affect the chosen process.

Out of the different channels of Higgs production, the $Vh~ (V = W, Z)$ production cross-section turn out to be more sensitive to the variation of effective operators concerned in the modification of the aforementioned couplings. In this analysis, we study the $Zh$ channel. There have been a number of analysis concerning the $Wh/Zh$ production modes \cite{Banerjee:2018bio, Banerjee:2019pks, Banerjee:2019twi, Freitas:2019hbk, Bishara:2022vsc,Bishara:2020pfx,Englert:2024ufr}. However, the effect of dipole contact operators in context of the $qqZ/qqZh$ couplings have been neglected. The effect of dipole operators have been studied in \cite{Bonnefoy:2024gca}, but concerning the $Wh$ mode only. We show that the presence of both Higgs-current as well as dipole contact operators appreciably alter the existing bounds on the concerned couplings at the present and projected LHC sensitivities.

The organization of the paper is as follows: we discuss SMEFT operators relevant for our analysis and their sensitivities in section \ref{s2}, constraints on the coupling modifiers as well as SMEFT coefficients from current data in section \ref{s3}, detailed collider analysis in section \ref{s4} and concluding remarks in \ref{s5}.

\section{EFT framework} 
\label{s2}
The couplings of our interest for this study are the $qqZ/qqZh$ and $hZZ$ vertices which appear in both VBF as well as $Zh$ production modes of the Higgs boson. 
The set of interaction terms involve those connecting to the gauge fields, field strength tensor (or the dual of it) along with the Higgs field and quark bilinear with all possible 
Lorentz structure, adhering to $U(1)_{\rm EM}$ gauge symmetry after EWSB, as furnished below.
	
\begin{align} \label{f00}
	\mathcal{L}_{SM} & = g_{uL}^{\tt SM} \; Z_{\mu} \overline{u_{L}} \gamma^{\mu} u_{L} + g_{uR}^{\tt SM} \; Z_{\mu} \overline{u_{R}} \gamma^{\mu} u_{R} + g_{dL}^{\tt SM} \; Z_{\mu} \overline{d_{L}} \gamma^{\mu} d_{L} + g_{dR}^{\tt SM} \; Z_{\mu} \overline{d_{R}} \gamma^{\mu} d_{R}\,; \\
	 \label{f01} 
	\mathcal{L}_{\kappa} & = \kappa_{Z} \frac{m_{Z}^{2}}{v} \;Z_{\mu} Z^{\mu} h\,; \\
	 \label{f02}
	\mathcal{L}_{NP} \nonumber & =  g_{ZZ} \; \frac{h}{v} Z^{\mu \nu} Z_{\mu \nu} + g_{Z\widetilde{Z}} \; \frac{h}{v} Z^{\mu \nu} \widetilde{Z}_{\mu \nu} + g_{Z \gamma} \; \frac{h}{v} A^{\mu \nu} Z_{\mu \nu} + g_{\widetilde{Z} \gamma} \; \frac{h}{v} A^{\mu \nu} \widetilde{Z}_{\mu \nu} \\ \nonumber & + g_{u}^{L} \; Z_{\mu} \overline{u_{L}} \gamma^{\mu} u_{L} + g_{u}^{R} \; Z_{\mu} \overline{u_{R}} \gamma^{\mu} u_{R} + g_{d}^{L} \; Z_{\mu} \overline{d_{L}} \gamma^{\mu} d_{L} + g_{d}^{R} \; Z_{\mu} \overline{d_{R}} \gamma^{\mu} d_{R} \\ & + \delta_{u}^{L} \; \frac{h}{v} Z_{\mu} \overline{u_{L}} \gamma^{\mu} u_{L} + \delta_{u}^{R} \; \frac{h}{v} Z_{\mu} \overline{u_{R}} \gamma^{\mu} u_{R} + \delta_{d}^{L} \; \frac{h}{v} Z_{\mu} \overline{d_{L}} \gamma^{\mu} d_{L} + \delta_{d}^{R} \; \frac{h}{v} Z_{\mu} \overline{d_{R}} \gamma^{\mu} d_{R} \\ \nonumber & + g_{u} \; \frac{1}{v} (\overline{u_{L}} \sigma^{\mu \nu} u_{R} + \overline{u_{R}} \sigma^{\mu \nu} u_{L}) Z_{\mu \nu} + g_{d} \; \frac{1}{v} (\overline{d_{L}} \sigma^{\mu \nu} d_{R} + \overline{d_{R}} \sigma^{\mu \nu} d_{L}) Z_{\mu \nu} \\ \nonumber & + \delta_{u} \; \frac{h}{v^{2}} (\overline{u_{L}} \sigma^{\mu \nu} u_{R} + \overline{u_{R}} \sigma^{\mu \nu} u_{L}) Z_{\mu \nu} + \delta_{d} \; \frac{h}{v^{2}} (\overline{d_{L}} \sigma^{\mu \nu} d_{R} + \overline{d_{R}} \sigma^{\mu \nu} d_{L}) Z_{\mu \nu}\,.
\end{align}
Note here in writing Eq. (\ref{f00}$-$\ref{f02}), we have adhered to $H=\left(\begin{array}{c} 0\\ \frac{v+h}{\sqrt{2}}\end{array}\right)$ after EWSB. Here, $\kappa_{Z}$ is the SM $hZZ$ coupling modifier. The advantage of $\kappa$ framework in addition to heavy integrable NP scenarios, it enables us to consider undetected light degrees of freedom without explicitly introducing a NP scale. In the above equations, the notations have their usual meanings, particularly note $Z_{\mu\nu}=\partial_\mu Z_\nu-\partial_\nu Z_\mu$, $ \widetilde{Z}_{\mu \nu}=\epsilon_{\mu\nu\alpha\beta}Z^{\alpha\beta}$ where $Z_\mu$ denotes the physical $Z$ boson, $\epsilon_{\mu\nu\alpha\beta}$ is a completely antisymmetric tensor and {$\sigma^{\mu\nu}= \frac{i}{2} [\gamma^\mu, \gamma^\nu]$}.

In absence of a direct observational evidence of NP signal at current experiments, one of the best possible ways to gauge the effects of NP is through 
higher dimensional effective operators constructed out of the SM fields respecting the SM gauge symmetry. 
They are popularly referred as SMEFT operators. Note here that SMEFT operators also include the Higgs field 
as isodoublet under the SM gauge group, while there are efforts to keep it beyond, for example, in HEFT \cite{Alonso:2015fsp}, 
but given the Higgs properties closely mimicking that of a SM doublet, we choose SMEFT framework to explore the limit of 
NP\footnote{Operators with two Higgs doublets \cite{Crivellin:2016ihg,Anisha:2019nzx}, or a right handed neutrino \cite{delAguila:2008ir,Aparici:2009fh,Bhattacharya:2015vja,Liao:2016qyd} have also been found out.}. 
These operators are suppressed by an appropriate power of a heavy NP mass scale ($\Lambda$) integrated out, 
depending on the mass-scale of the operator. Understandably the effects of such operators diminish with higher mass dimension of the operators, 
hence we will constrict ourselves to the lowest dimension operators that contribute to the $Zh$ associated production, which appears in dimension 
6\footnote{It is well known that the only SMEFT operator in dimension 5 is the Weinberg operator $\bar{N}N^c;~N=H^\dagger \epsilon \ell$, which generates Majorana neutrino mass.}; operators of dimension 8 \cite{Li:2020gnx,Murphy:2020rsh} may also contribute, but we have not considered them in our analysis. We will provide a detailed 
matching of Eq.\ref{f02} with the standard SMEFT framework in Warsaw basis \cite{Grzadkowski:2010es} later.


\subsection{SMEFT operators relevant to $Zh$ production}

The SMEFT Lagrangian can be written as:
\begin{equation}
	\mathcal{L} = \mathcal{L}_{SM} + \sum_{i,(n>4)} \frac{c_{i}}{\Lambda^{(n-4)}}\mathcal{O}_{i}^{(n)}\,,
\end{equation}
Here, $\mathcal{O}_{i}^{(n)}$ represents operators of dimension $n$ and $c_{i}$ represents dimensionless couplings or Wilson's coefficients. Throughout this analysis, the effective scale is taken as $\Lambda = 1$ TeV, while the resulting bound or the sensitivity is obtained for the corresponding Wilson coefficient $c_{i}$. However, it is simple to understand that the limit on $c_i$ can be rescaled easily for different choices of $\Lambda$, following, $\frac{c_i}{\Lambda^2}=\frac{c'_i}{{\Lambda'}^2}$. Although, the validity of effective limit depends on the choice of $\Lambda$, strictly speaking the center of mass (CoM) energy of the reaction should abide by $\sqrt{s}<\Lambda$. At hadron collider, it is difficult to ensure such a condition, we elaborate upon the possibility later.

\begin{figure}[htbp]
	\begin{center}
		\begin{tikzpicture}
			\begin{feynman}
				\vertex (a);
				\vertex [above left = 1cm of a] (b) {$q$};
				\vertex [below left = 1cm of a] (c) {$\Bar{q}$};
				\vertex [right = 1cm of a] (d);
				\vertex [above right = 1cm of d] (e) {$Z$};
				\vertex [below right = 1cm of d] (f) {$h$};
		
				\diagram* {
					(b) -- [fermion] (a) -- [fermion] (c),
					(a) -- [boson, edge label = $Z$] (d),
					(d) -- [boson] (e),
					(d) -- [scalar] (f)
				};
			\end{feynman}
		\end{tikzpicture} \hspace{1cm}
		\begin{tikzpicture}
			\begin{feynman}
				\vertex (a);
				\vertex [above left = 1cm of a] (b) {$q$};
				\vertex [below left = 1cm of a] (c) {$\Bar{q}$};
				\vertex [right = 1cm of a, dot] (d) {};
				\vertex [above right = 1.5cm of d] (e) {$Z$};
				\vertex [below right = 1.5cm of d] (f) {$h$};
				
				\diagram* {
					(b) -- [fermion] (a) -- [fermion] (c),
					(a) -- [boson, edge label = $Z/ \gamma$] (d),
					(d) -- [boson] (e),
					(d) -- [scalar] (f)
				};
			\end{feynman}
		\end{tikzpicture}
		\begin{tikzpicture}
			\begin{feynman}
				\vertex [dot] (a) {};
				\vertex [above left = 1.5cm of a] (b) {$q$};
				\vertex [below left = 1.5cm of a] (c) {$\Bar{q}$};
				\vertex [above right = 1.5cm of a] (e) {$Z$};
				\vertex [below right = 1.5cm of a] (f) {$h$};
				
				\diagram* {
					(b) -- [fermion] (a) -- [fermion] (c),
					(a) -- [boson] (e),
					(a) -- [scalar] (f)
				};
			\end{feynman}
		\end{tikzpicture}
		\begin{tikzpicture}
			\begin{feynman}
				\vertex [dot] (a) {};
				\vertex [above left = 1.5cm of a] (b) {$q$};
				\vertex [below left = 1.5cm of a] (c) {$\Bar{q}$};
				\vertex [right = 1cm of a] (d);
				\vertex [above right = 1cm of d] (e) {$Z$};
				\vertex [below right = 1cm of d] (f) {$h$};
				
				\diagram* {
					(b) -- [fermion] (a) -- [fermion] (c),
					(a) -- [boson, edge label = $Z$] (d),
					(d) -- [boson] (e),
					(d) -- [scalar] (f)
				};
			\end{feynman}
		\end{tikzpicture}
		\caption{SM and EFT contributions to the $Zh$ associated production mode at the LHC. Dotted vertices refer to the presence of EFT operators. \label{fig:3}}
	\end{center}
\end{figure}

There is no unique basis for the choice of these operators and all the different bases are equivalent. We choose Warsaw basis for our representation. The SMEFT operators relevant to $Zh$ associated production mode are tabulated in Table \ref{tab:1} and the corresponding 
Feynman diagrams are shown in Figure \ref{fig:3}. The vertices with a blob represent effective operator contribution, which modifies $hZZ$, $qqZ$ vertices and additional contributions are obtained via $\gamma Zh$ and $qqZh$ vertices.

Before moving forward, it should be noted that operators are classified majorly into two categories, those generated at loop level (LG operators) or potential tree generated 
(PTG operators) for NPs those which are weakly coupled and decoupling \cite{Bhattacharya:2021edh}. The LG operators are suppressed additionally by a factor of 
$(4 \pi)^{2}$ coming from the loops and hence their contributions are expected to be suppressed in comparison to the PTG operators if the underlying NP scale is the same. However, this additional suppression factor for the LG operators is not intrinsically incorporated in most SMEFT studies and in order to maintain uniformity of the nomenclature, we will however use the operator classification as in \cite{Grzadkowski:2010es}.

\begin{table}[htbp]
	\begin{center}
		\begin{multicols}{2}
			{\renewcommand{\arraystretch}{1.1}
			\begin{tabular}{|l|}
				\hline
				\textbf{CP even $hZZ$ vertex modifiers} \\ \hline
				$\mathcal{O}_{H \square} = (H^{\dagger} H) \square (H^{\dagger} H)$ \\
				$\mathcal{O}_{HDD} = (H^{\dagger} D_{\mu} H)(H^{\dagger} D^{\mu} H)$ \\
				$\mathcal{O}_{HW} = (H^{\dagger} H) W^{i}_{\mu \nu} W^{i \mu \nu}$ \\
				$\mathcal{O}_{HB} = (H^{\dagger} H) B_{\mu \nu} B^{\mu \nu}$ \\
				$\mathcal{O}_{HWB} = (H^{\dagger} \tau^{i} H) W^{i}_{\mu \nu} B^{\mu \nu}$ \\
				\hline
				\textbf{CP odd $hZZ$ vertex modifiers} \\ \hline
				$\mathcal{O}_{H\widetilde{W}} = (H^{\dagger} H) \widetilde{W}^{i}_{\mu \nu} W^{i \mu \nu}$ \\
				$\mathcal{O}_{H\widetilde{B}} = (H^{\dagger} H) \widetilde{B}_{\mu \nu} B^{\mu \nu}$ \\
				$\mathcal{O}_{H\widetilde{W}B} = (H^{\dagger} \tau^{i} H) \widetilde{W}^{i}_{\mu \nu} B^{\mu \nu}$ \\				
				\hline
			\end{tabular}}
			\columnbreak
			
			\begin{tabular}{|l|}
			\hline
			\textbf{Higgs-current operators} \\ \hline
			$\mathcal{O}_{Hq}^{(1)} = (H^{\dagger} i \overleftrightarrow{D_{\mu}} H)(\Bar{q}_{p} \gamma^{\mu} q_{r})$ \\
			$\mathcal{O}_{Hq}^{(3)} = (H^{\dagger} i \overleftrightarrow{D_{\mu}^{i}} H)(\Bar{q}_{p} \tau^{i} \gamma^{\mu} q_{r})$ \\
			$\mathcal{O}_{Hu} = (H^{\dagger} i \overleftrightarrow{D_{\mu}} H)(\Bar{u}_{p} \gamma^{\mu} u_{r})$ \\
			$\mathcal{O}_{Hd} = (H^{\dagger} i \overleftrightarrow{D_{\mu}} H)(\Bar{d}_{p} \gamma^{\mu} d_{r})$ \\ 
			\hline
			\textbf{Dipole operators} \\ \hline
			$\mathcal{O}_{uW} = (\Bar{q}_{p} \sigma^{\mu \nu} u_{r}) \tau^{i} \widetilde{H} W^{i}_{\mu \nu}$ \\
			$\mathcal{O}_{uB} = (\Bar{q}_{p} \sigma^{\mu \nu} u_{r}) \widetilde{H} B_{\mu \nu}$ \\
			$\mathcal{O}_{dW} = (\Bar{q}_{p} \sigma^{\mu \nu} d_{r}) \tau^{i} H W^{i}_{\mu \nu}$ \\
			$\mathcal{O}_{dB} = (\Bar{q}_{p} \sigma^{\mu \nu} d_{r}) H B_{\mu \nu}$ \\ \hline
			\end{tabular}			
		\end{multicols}
	\end{center}
	\caption{Dimension 6 SMEFT operators: \textit{Left:} $hZZ$ vertex modifiers, \textit{Right:} $qqZ/qqZh$ contact vertices (Higgs-current and dipole types), relevant to $Vh$ associated production at the LHC. Here, $p$ and $r$ are quark family indices. For complex operators, the presence of respective hermitian conjugates is assumed. For relevance: $D_{\mu} = \partial_{\mu} + i g \frac{\tau^{i}}{2} W^{i}_{\mu} + i g' \frac{1}{2} B_{\mu}$, $W_{\mu \nu}^{i} = \partial_{\mu} W_{\nu}^{i} - \partial_{\nu} W_{\mu}^{i} + g \epsilon^{ijk} W_{\mu}^{j} W_{\nu}^{k}$, $B_{\mu \nu} = \partial_{\mu} B_{\nu} - \partial_{\nu} B_{\mu}$, $ H^{\dagger} i \overset{\leftrightarrow}{D_{\mu}} H= i H^{\dagger} D_{\mu} H - i (D_{\mu} H)^{\dagger} H$, $H^{\dagger} i \overset{\leftrightarrow}{D_{\mu}^{i}} H = i H^{\dagger} \tau^{i} D_{\mu} H - i (D_{\mu} H)^{\dagger} \tau^{i} H$, $\widetilde{V}_{\mu \nu} = \epsilon_{\mu \nu \rho \sigma} V^{\rho \sigma}$ ($V = W^{i}, B$). \label{tab:1}}
\end{table}

A few observations related to the operators as mentioned in Table \ref{tab:1} are as follows: The operators $\mathcal{O}_{H \square}$ and $\mathcal{O}_{HDD}$ modify the Higgs gauge boson coupling by a multiplicative factor without introducing any new Lorentz structure, amounting to a renormalization of the Higgs field. As such the contribution can be absorbed in the coupling modifier. The operators $\mathcal{O}_{HW}$, $\mathcal{O}_{HB}$ and $\mathcal{O}_{HWB}$ contribute in the modification of the $hZZ$ vertex. These amounts to the CP conserving anomalous Higgs gauge boson coupling. The operators $\mathcal{O}_{H\widetilde{W}}$, $\mathcal{O}_{H\widetilde{B}}$ and $\mathcal{O}_{H\widetilde{W}B}$ contribute in the modification of the $hZZ$ vertex. These amounts to the CP violating anomalous Higgs gauge boson coupling. The operators $\mathcal{O}^{(1)}_{Hq}$, $\mathcal{O}^{(3)}_{Hq}$, $\mathcal{O}_{Hu}$ and $\mathcal{O}_{Hd}$ contribute to the 4-point interaction $qqZh$. These operators also contribute to $qqZ$ vertex. These operators are referred to as Higgs-current operators since their form is analogous to current, $j^{\mu} = \overline{\psi} \gamma^{\mu} \psi$. The operators $\mathcal{O}_{uW}$, $\mathcal{O}_{uB}$, $\mathcal{O}_{dW}$ and $\mathcal{O}_{dB}$ contribute to the 4-point interaction $qqZh$. These operators also contribute to $qqZ$ vertex. These operators contribute to the dipole moments of quarks and hence can be referred to as dipole operators. A complete matching between the coefficients of $\mathcal{L}_{NP}$ in Eq.\ref{f00} and the SMEFT coefficients as in Table~\ref{tab:1} are shown in Table \ref{tab01}.

\begin{table}[htbp]
	\centering
	\begin{tabular}{|c|c|}
		\hline
		Coefficient & Warsaw equivalent \\ \hline
		$g_{ZZ}$ & $2\left(c_{w}^{2} c_{HW} + s_{w}^{2} c_{HB} + s_{w} c_{w} c_{HWB} \right)v^{2}/ \Lambda^{2} $ \\ 
		$g_{Z \gamma}$ & $\left(2 s_{w} c_{w}^{2} (c_{HW} - c_{HB}) + (s_{w}^{2} - c_{w}^{2}) c_{HWB} \right)v^{2}/ \Lambda^{2}$ \\ 
		$g_{Z \widetilde{Z}}$ & $2 \left(c_{w}^{2} c_{H\widetilde{W}} + s_{w}^{2} c_{H\widetilde{B}} + s_{w} c_{w} c_{H\widetilde{W}B} \right)v^{2}/ \Lambda^{2}$ \\ 
		$g_{\widetilde{Z} \gamma}$ & $ \left(2 s_{w} c_{w}^{2} (c_{H\widetilde{W}} - c_{H\widetilde{B}}) + (s_{w}^{2} - c_{w}^{2}) c_{H\widetilde{W}B} \right)v^{2}/ \Lambda^{2}$ \\ 
		$g^{L}_{u}$ & $- \left(c_{Hq}^{(1)} + c_{Hq}^{(3)}\right) ev^{2}/ \left(2\Lambda^{2} s_{w} c_{w}\right)$ \\ 
		$g^{L}_{d}$ & $- \left(c_{Hq}^{(1)} - c_{Hq}^{(3)}\right) ev^{2}/ \left(2\Lambda^{2} s_{w} c_{w}\right)$ \\ 
		$g^{R}_{u}$ & $- c_{Hu}ev^{2}/ \left(2\Lambda^{2} s_{w} c_{w}\right)$ \\ 
		$g^{R}_{d}$ & $- c_{Hd} ev^{2}/ \left(2\Lambda^{2} s_{w} c_{w}\right)$ \\ 
		$\delta^{L}_{u}$ & $- \left(c_{Hq}^{(1)} + c_{Hq}^{(3)}\right) ev^{2}/ \left(\Lambda^{2} s_{w} c_{w}\right)$ \\ 
		$\delta^{L}_{d}$ & $- \left(c_{Hq}^{(1)} - c_{Hq}^{(3)}\right) ev^{2}/ \left(\Lambda^{2} s_{w} c_{w}\right)$ \\ 
		$\delta^{R}_{u}$ & $- c_{Hu}ev^{2}/ \left(\Lambda^{2} s_{w} c_{w}\right)$ \\ 
		$\delta^{R}_{d}$ & $- c_{Hd} ev^{2}/ \left(\Lambda^{2} s_{w} c_{w}\right)$ \\ 
		$g_{u}$ & $ \left(s_{w} c_{uB} - c_{w} c_{uW}\right)ev^{2}/ \sqrt{2} \Lambda^{2}$ \\ 
		$g_{d}$ & $ \left(s_{w} c_{dB} + c_{w} c_{dW}\right) ev^{2}/ \sqrt{2} \Lambda^{2}$ \\ 
		$\delta_{u}$ & $ \sqrt{2} \left(s_{w} c_{uB} - c_{w} c_{uW}\right)ev^{2}/ \Lambda^{2}$ \\ 
		$\delta_{d}$ & $ \sqrt{2} \left(s_{w} c_{dB} + c_{w} c_{dW}\right) ev^{2}/ \Lambda^{2}$ \\ \hline		
	\end{tabular}
	\caption{Matching between coefficients in equation (\ref{f02}) and the SMEFT coefficients. $s_{w}$ and $c_{w}$ are the sine and cosine of the Weinberg angle. \label{tab01}}
\end{table}


\subsection{Parametrization of production cross sections}

\begin{figure}[htbp]
	\includegraphics[trim=0 0 1.5cm 1.25cm, clip, width = 0.475 \textwidth]{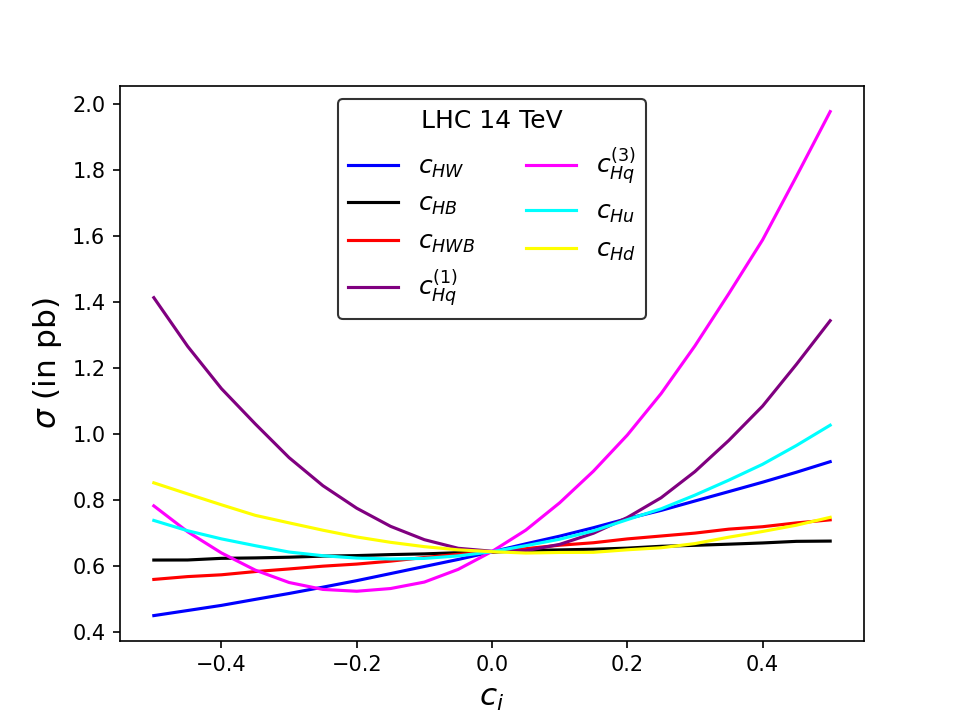}
	\hfill
	\includegraphics[trim=0 0 1.5cm 1.25cm, clip, width = 0.475 \textwidth]{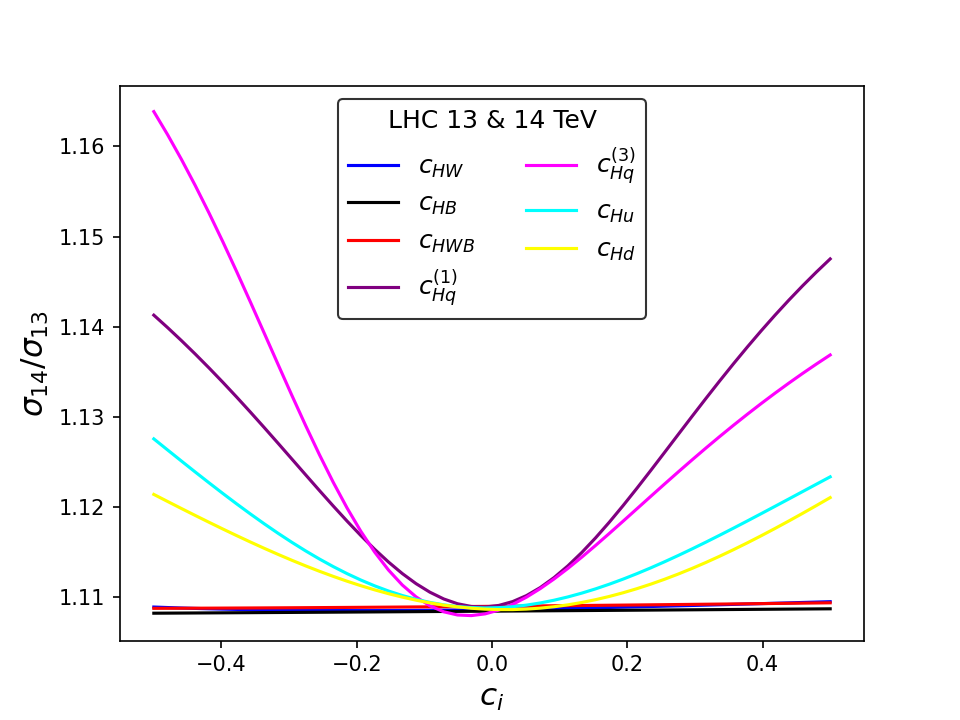}
	\hfill
	\caption{\textit{Left:} Variation of $Zh$ associated production cross section with variation in CP even $hZZ$ modifiers and Higgs-current effective operator coefficients at the 14 TeV LHC. \textit{Right:} Variation in the ratio of production cross section of the 14 TeV LHC to the 13 TeV LHC with variation in the effective operator coefficients. See figure inset for the colour codes. 
	\label{fig:4}}
\end{figure}

\begin{figure}[htbp]
	\includegraphics[trim=0 0 1.5cm 1.25cm, clip, width = 0.45 \textwidth]{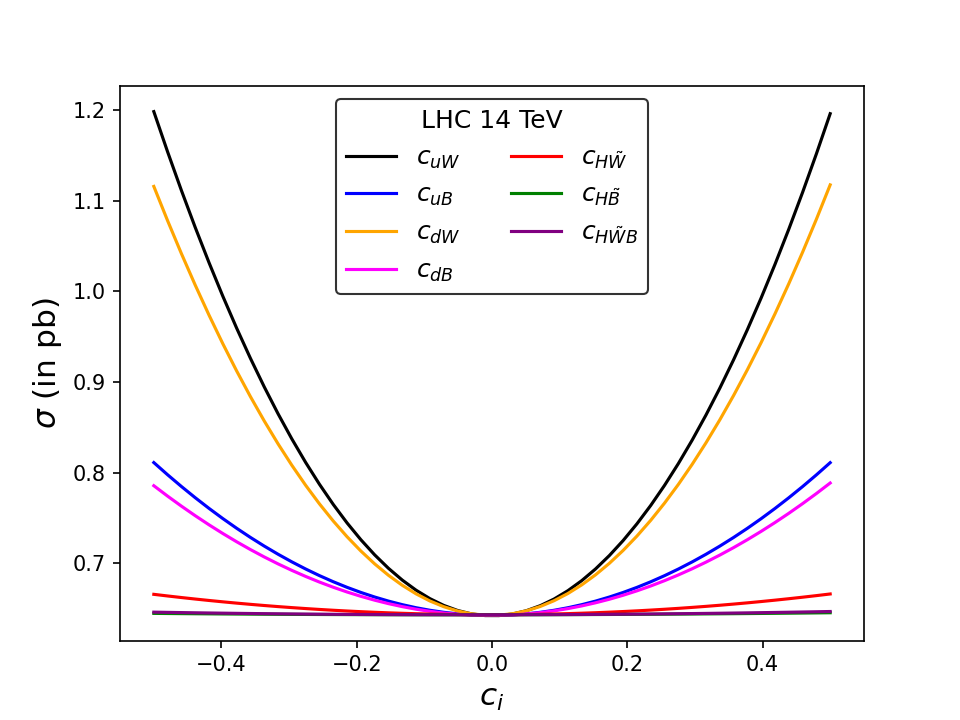}
	\hfill
	\includegraphics[trim=0 0 1.5cm 1.25cm, clip, width = 0.45 \textwidth]{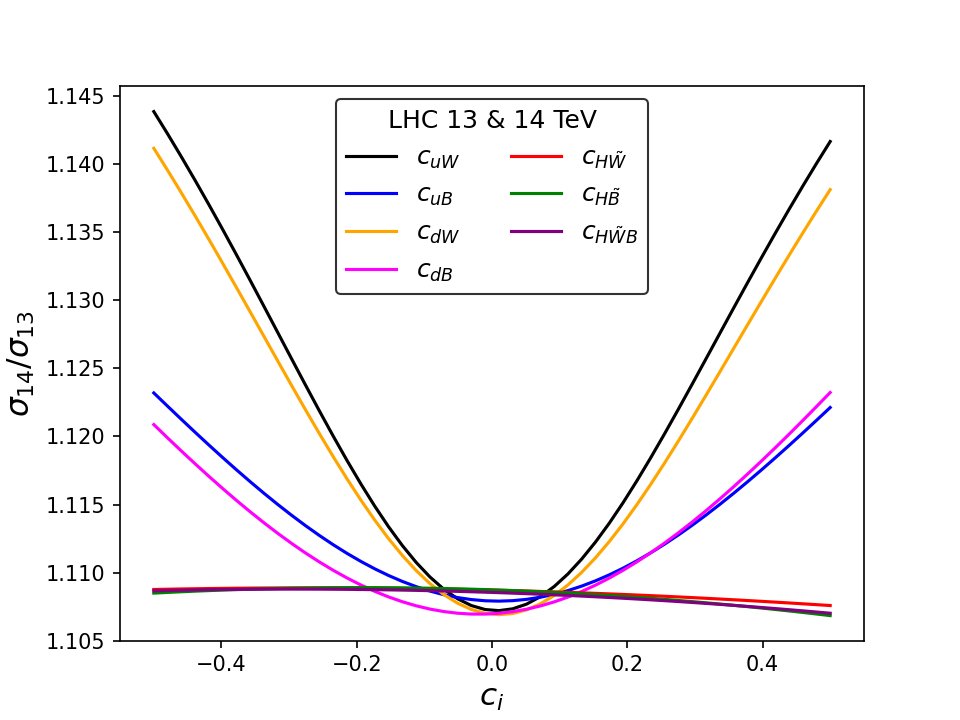}
	\hfill
	\caption{\textit{Left:} Variation of $Zh$ associated production cross section with change in CP odd $hZZ$ modifiers and dipole effective operator coefficients at the 14 TeV LHC. \textit{Right:} Variation in the ratio of production cross section of the 14 TeV LHC to the 13 TeV LHC with change in effective operator coefficients. See figure inset for the colour codes. 
	\label{fig:5}}
\end{figure}

When SMEFT effects are included, the cross section consists of three types of terms viz. the pure SM contribution, the SM plus EFT interference ($\propto \Lambda^{-2}$ from dimension six operators) and the pure EFT contribution ($\propto \Lambda^{-4}$). Inclusion of dimension 8 operators would result in SM plus EFT interference contribution $\propto \Lambda^{-4}$ which is comparable to the pure EFT contribution from dimension 6 terms, however, we do not consider them in this analysis. The production cross section in presence of SMEFT operators takes the form:
\begin{equation} \label{eq12}
	\sigma = \sigma^{(SM)} + \sum_{i=1}^{n_{L}}c_i\sigma^{(L)}_{i} +
	\sum_{i \le j}^{n_{Q}}c_i c_j \sigma^{(Q)}_{ij}\,,
\end{equation}
where, $n_L$ and $n_Q$ refers to the number of linear and quadratic SMEFT contributions. The effective scales $\Lambda^{-2}$ and $\Lambda^{-4}$ are absorbed in the cross sections $\sigma^{(L)}_{i}$ and $\sigma^{(Q)}_{ij}$ respectively. In presence of coupling modifiers $\kappa$, the cross section in equation (\ref{eq12}) takes the form as below,
\begin{equation} \label{eq11}
	\begin{split}
	\sigma &= \kappa^2 \; \sigma^{(SM)} + \sum_{i=1}^{n_{L}} \; (\alpha^{(L)}_{i} \kappa + \beta^{(L)}_{i} \kappa^{2})\; c_i\;\sigma^{(L)}_{i} +
	\sum_{i \le j}^{n_{Q}} \; (1 + \alpha^{(Q)}_{ij} \kappa + \beta^{(Q)}_{ij} \kappa^{2})\;  c_i c_j\; \sigma^{(Q)}_{ij}\,, \\
	\mu &= \kappa^2 + \sum_{i=1}^{n_{L}} \; (\alpha^{(L)}_{i} \kappa + \beta^{(L)}_{i} \kappa^{2})\; c_i\;\mu^{(L)}_{i} +
	\sum_{i \le j}^{n_{Q}} \; (1 + \alpha^{(Q)}_{ij} \kappa + \beta^{(Q)}_{ij} \kappa^{2})\;  c_i c_j\; \mu^{(Q)}_{ij}\,, \\	
	\end{split}
\end{equation}
where, $\mu$ is the signal strength, i.e. $\mu = \sigma/\sigma_{SM}$. Differential cross sections can be parametrized in a similar manner. At the hadron collider, an analytical form of the cross section is difficult to write due to the uncertainty of the incoming momenta of the partons. However, we can numerically fit the production cross section as a function of the parameters and this turns out to be equally useful for gauging the effects of NP. In the following, we fit the operator coefficients following \eqref{eq11} at 14 TeV LHC for the $Zh$ associated production process considering one operator at a time. The variation of the total production cross section are shown in Figure \ref{fig:4} (for CP even $hZZ$ modifiers and Higgs-current operators) and \ref{fig:5} (for CP odd $hZZ$ modifiers and dipole operators). The ratio of the total production cross section for moving from 13 TeV to 14 TeV are also shown. We adopt the following methodology, the effective Lagrangian is implemented in \texttt{FeynRules} \cite{Alloul:2013bka} to generate the \texttt{UFO} model \cite{Degrande:2011ua} file. The model file is validated with standard UFOs like \texttt{SMEFTsim} \cite{Brivio:2020onw} and \texttt{SMEFTatNLO} \cite{Degrande:2020evl}. This model file is then fed into the event generator \texttt{MG5\_aMC} \cite{Alwall:2011uj} to obtain the above cross sections at the leading order (LO). The cross sections are generated at production level without decaying $Z$ or $h$ to any specific decay mode without any kinematic cuts. The PDF set and flavor scheme choice are $\mathtt{NN23LO1}$ PDF set and $U(2)_{q} \times U(2)_{u} \times U(2)_{d}$, respectively. The NP scale is chosen as $\Lambda =$ 1 TeV. A few key points observed are as follows: The Higgs-current operators have significant linear as well as quadratic contributions indicating large pure EFT contribution. The enhancement of the EFT terms result from the absence of propagator suppression in the production process. The dipole operators have no linear contribution. This is because they bind fermions with different chiralities and hence do not interfere with the SM process. Due to the absence of linear dependence, they cause variation in the cross section only in the positive direction. All the operators except $\mathcal{O}_{Hq}^{(3)}$ shows near symmetric behaviour along the positive and negative side. The operator $\mathcal{O}_{Hq}^{(3)}$ dips around $c_{Hq}^{(3)} = -0.2$ and then continues to rise as we go further in the negative direction. This is due to strong linear as well as quadratic dependence	and the trade off between them happens away from $c_{Hq}^{(3)} = 0$ towards the negative axis. When moving from 13 TeV to 14 TeV LHC, the contact operators i.e. operators associated with $qqZ/qqZh$ vertices, significantly boost the production cross sections. The other operators are less sensitive.

For our analysis, we consider the two operators $\mathcal{O}^{(3)}_{Hq}$ and $\mathcal{O}_{uW}$ as a representative of the Higgs-current and dipole class respectively. It should be noted that the choice is based on their sensitivity to the production cross section, however other operators from the respective classes maybe more sensitive to particular observables or be less bounded from experiments. Our consideration is based on proof of concept and any other operator can be chosen in lieu of these. Along with the coupling modifier $\kappa_{Z}$ (which approximately estimates the EFT effect to the $hZZ$ vertex), this form our EFT parameter space: \{$\kappa_{Z}$, $c_{uW}$, $c_{Hq}^{(3)}$\}. The parametrized coefficients for signal strength associated with $Zh$ production with these operators following Eqn. \eqref{eq11} are tabulated in Table \ref{tab:par}. This parametrization gives an overview of the EFT effects on the inclusive cross section. Coefficients of linear terms in $1$, $\kappa_{Z}$, $c_{uW}$ and $c_{Hq}^{(3)}$ are not present in our parametrization as it is evident from Eq. (\ref{eq11}) that the interference between SM and SMEFT vertices appear as coefficient of $\kappa_{Z}\;c^{(3)}_{Hq}$. Further, the coefficients of $\kappa_{Z}\; c_{uW}$, $\kappa_{Z}^{2}\; c_{uW}$, $c_{uW}\; c_{Hq}^{(3)}$, $\kappa_{Z}\; c_{uW}\; c_{Hq}^{(3)}$ and $\kappa_{Z}^{2}\; c_{uW}\; c_{Hq}^{(3)}$ vanish because of no interference between the SM and $\mathcal{O}_{Hq}^{(3)}$ operators with diagrams with $\mathcal{O}_{uW}$ operator as the dipole operator mixes quarks of different chiralities ($\overline{q}_{L} \sigma_{\mu \nu} q_{R} \;/\; \overline{q}_{R}  \sigma_{\mu \nu} q_{L}$) in contrary to the SM or Higgs-current operators which mixes quarks of same chirality ($\overline{q}_{L} \gamma_{\mu} q_{L} \;/\; \overline{q}_{R} \gamma_{\mu} q_{R}$).

We shall also note that the variation of the cross-section presented here with EFT operator coefficients, have been done without considering the constraints on these operators. We shall do it in the next section. We get the dominant physics contribution to the process under consideration and the resultant parametrisation, which remains intact with the modified range of the operator coefficients that will be used for scanning the parameter space.

\begin{center}
	\begin{table}[!htbp]
		\centering
		\begin{tabular}{|c|c|c|c|c|c|c|}
			\hline
			 Coefficients & $1 \times 1$ & $1 \times c_{uW}$ & $1 \times c_{Hq}^{(3)}$ &  $c_{uW} \times c_{uW}$ & $c_{Hq}^{(3)} \times c_{Hq}^{(3)}$ & $c_{uW} \times c_{Hq}^{(3)}$ \\ \hline
			 $1 \times 1$ & $-$ & $-$ & $-$ & 3.0277 & 4.2429 & $-$  \\ \hline
			 $1 \times \kappa_{Z}$ & $-$ & $-$ & 1.7060 & 0.2106 & 0.3880 & $-$  \\ \hline
			$\kappa_{Z} \times \kappa_{Z}$ & 1.0000 & $-$ & 0.1439 & 0.2000 & $-$0.0472 & $-$ \\  \hline
		\end{tabular}
		\caption{The parametrized coefficients ($\mu_{i}$, $\mu_{ij}$) for operators $\mathcal{O}^{(3)}_{Hq}$, $\mathcal{O}_{uW}$ and $\kappa_{Z}$ contributing 
		to $Zh$ production at 14 TeV LHC, following equation (\ref{eq11}).}
		\label{tab:par}
	\end{table}
\end{center}

\begin{figure}[htbp]
	\includegraphics[trim=0 0 1.125cm 1.25cm, clip, width = 0.475 \textwidth]{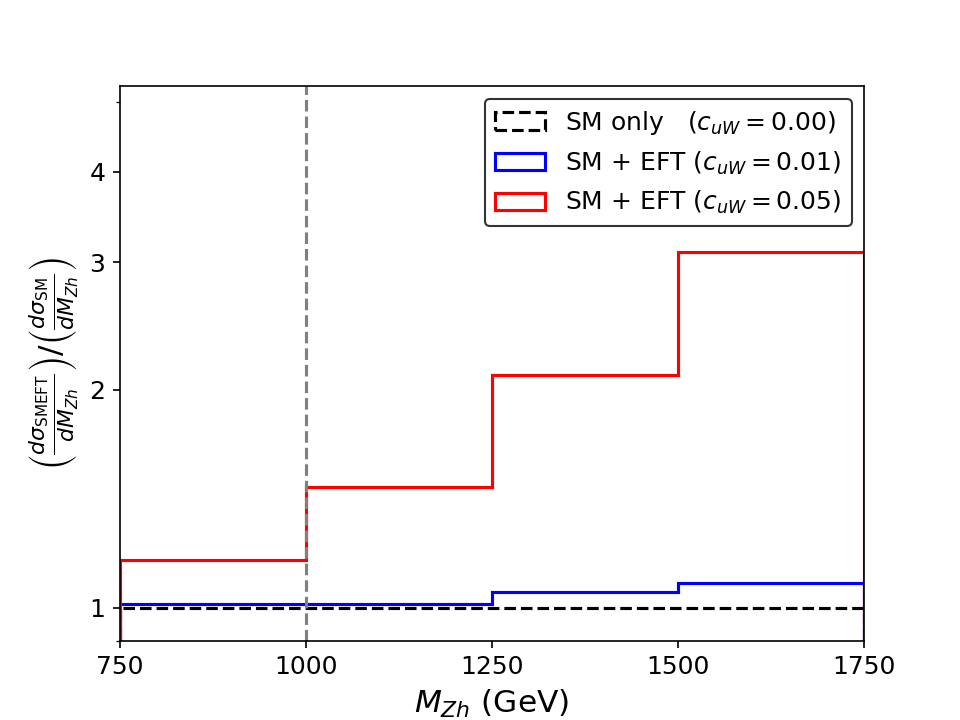}
	\hfill
	\includegraphics[trim=0 0 1.125cm 1.25cm, clip, width = 0.475 \textwidth]{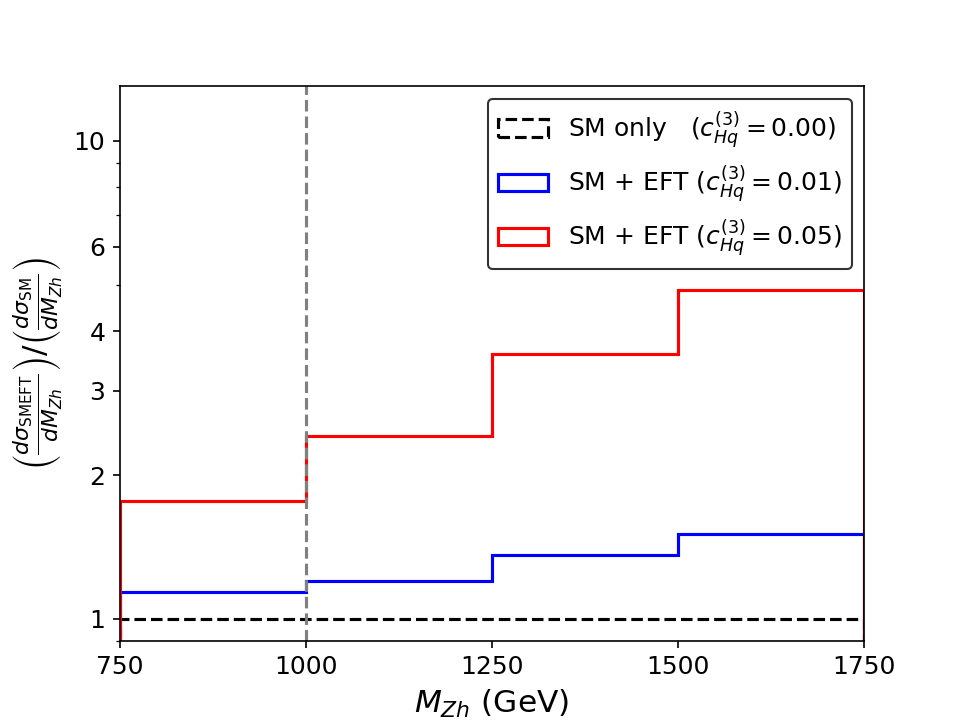}
	\hfill
	\caption{\textit{Left:} Ratio of the invariant mass distribution, $M_{Zh}$ for $Zh$ production in presence of $\mathcal{O}_{uW}$ operator to the corresponding SM prediction at the LHC 14 TeV. \textit{Right:} Same for $\mathcal{O}^{(3)}_{Hq}$ operator. The grey vertical line refers to the EFT scale, $\Lambda = 1$ TeV. \label{fig:7a}}
\end{figure}

It should be further noted that due to the presence of momentum dependence in EFT vertices or absence of propagators, effect of EFT operators are evident mostly at the tails of differential distribution. Hence, presence of EFT, if any, can be best interpreted from changes to the differential distributions. Figure \ref{fig:7a} shows the ratio of the invariant mass distribution, $M_{Zh} = \sqrt{\left(p_{Z} + p_{h}\right)^{2}}$ for $Zh$ production in the presence of EFT operators to the corresponding SM prediction at the 14 TeV LHC.

\subsection{Comparison with other Higgs production modes}

\begin{figure}[htbp]
	\includegraphics[trim=0 0 0.25cm 0.25cm, clip, width = 0.475 \textwidth]{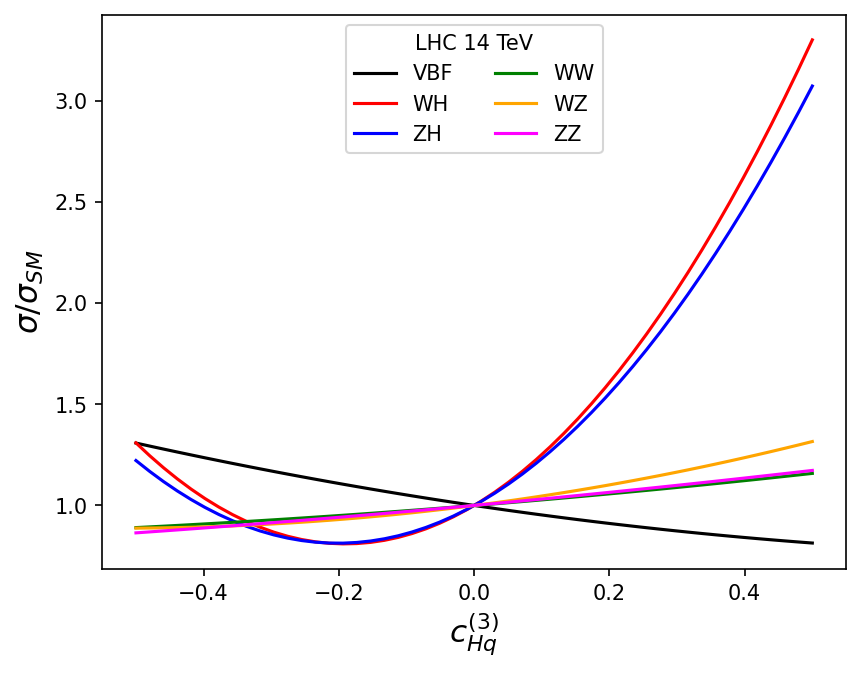}
	\hfill
	\includegraphics[trim=0 0 0.25cm 0.25cm, clip, width = 0.475 \textwidth]{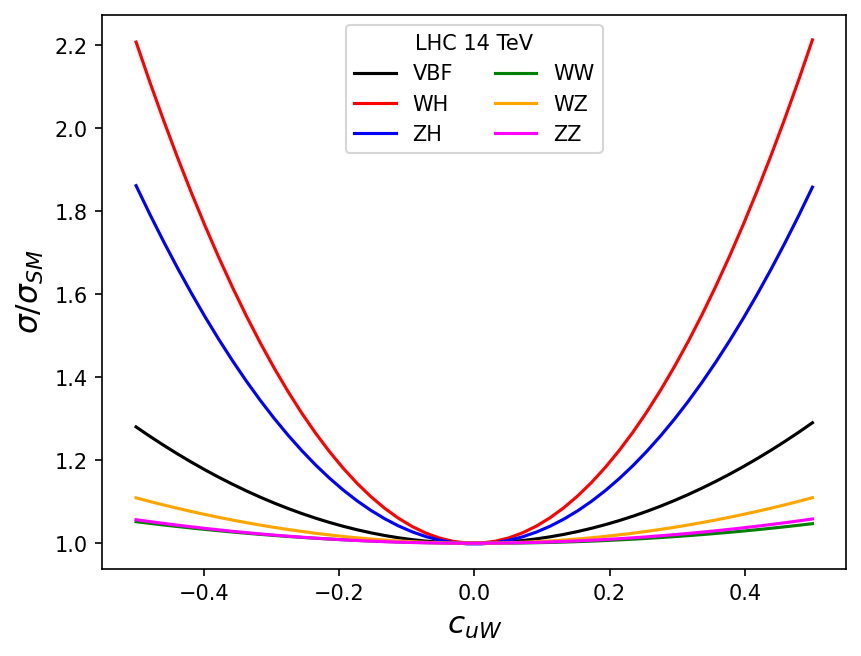}
	\hfill
	\caption{\textit{Left:} Variation in the ratio of production cross section to the SM cross section in VBF, $Vh$ and $VV$ processes at the 14 TeV LHC with 
	variation in effective coefficient $c_{Hq}^{(3)}$. \textit{Right:} Same for $c_{uW}$. \label{fig:7}}
\end{figure}

The operators $\mathcal{O}_{uW}$ and $\mathcal{O}_{Hq}^{(3)}$ are not unique to the $Vh$ ($V = W, Z$) associated production mode. It appears in the vector boson fusion (VBF) and single top production in association with Higgs ($thj$) as well. The $thj$ process has a comparatively smaller cross section and it is not ideal for the study of these operators, but the VBF mode has a larger cross section compared to $Vh$ associated production mode, as considered here. These operators are also associated with the diboson production ($VV = WW, WZ, ZZ$) processes. However, the ratio of the signal production cross-section including EFT and that of the SM contribution is more sensitive to the changes in the 
Wilson coefficients in $Vh$ mode compared to that of VBF and $VV$ channels. This has been illustrated in Figure \ref{fig:7} for the operators $\mathcal{O}^{(3)}_{Hq}$ 
(left) and $\mathcal{O}_{uW}$ (right) \footnote{One should however note here that the sensitivity of a specific channel not only depends the signal cross-section, but also relies 
on the background contribution and the efficiency in extracting the signal.}. Also, for both $VV$ and $VBF$ modes, pure contact interaction diagrams are absent, so EFT contributions always come in presence 
of one or more propagators. Hence, for the study of $qqZ/qqZh$ vertices, $Zh$ production is considered in our analysis.

\section{Constraints on the $hZZ$ coupling from current data} \label{s3}

The $\kappa$ framework is the most preferred way to incorporate uncertainties in the measurement of Higgs couplings. This has been adopted in most of the experimental results reported so far, and we will also follow the same here. The current uncorrelated bounds from ATLAS (139 $fb^{-1}$) and CMS (138 $fb^{-1}$) are based on Run II data and involves global fit over the main five production and decay channels. Since we intend to study $Zh$ associated production with $h$ subsequently decaying to bottom pair, the bottom Higgs coupling modifier, $\kappa_{B}$ is also important for this analysis. Higgs decay to bottom pair has been studied intensely at the LHC across several channels and the bounds are given in terms of signal strength ($\mu$), i.e. the ratio of the observed signal to the SM predicted value. In Table \ref{tab:03} we list out the uncorrelated bounds on the coupling modifiers as well as the signal strengths of the Higgs decay to bottom pair in the $Zh$ production channel. 

\begin{table}[htbp]
	\centering
	\begin{tabular}{|c|c|c|}
		\hline
		Measurable & Bound & Source \\ \hline
		\multirow{2}{*}{$\kappa_{Z}$} & [0.97, 1.11] & CMS ($pp \rightarrow h$) \cite{CMS:2022dwd} \\
		& [0.94, 1.05] & ATLAS ($pp \rightarrow h$) \cite{ATLAS:2022vkf} \\ \hline
		\multirow{2}{*}{$\kappa_{B}$} & [0.85, 1.11] & CMS ($pp \rightarrow h$) \cite{CMS:2022dwd} \\
		& [0.79, 1.01] & ATLAS ($pp \rightarrow h$) \cite{ATLAS:2022vkf} \\ \hline
		\multirow{2}{*}{$\mu_{hbb}^{Zh}$} & [0.59, 1.17] & CMS ($pp \rightarrow Zh, h \rightarrow b \Bar{b}$) \cite{CMS:2018nsn} \\
		& [0.85, 1.33] & ATLAS ($pp \rightarrow Zh, h \rightarrow b \Bar{b}$) \cite{ATLAS:2020fcp} \\ \hline
	\end{tabular}
	\caption{Constraints on $\kappa_{Z}$, $\kappa_{B}$ and $\mu_{Hbb}^{Zh}$ (at 95\% C.L.) from ATLAS and CMS experiments. \label{tab:03}}
\end{table}

Considering the bounds on the signal strength, we can limit the coupling modifiers using the measured signal strengths as shown in Eqn. \ref{ee4}. 

\begin{equation} \label{ee4}
	\mu^{Zh}_{hbb} = \frac{(\sigma^{Zh} \times (B.R.)_{hbb})}{(\sigma^{Zh} \times (B.R.)_{hbb})_{SM}} = \frac{\kappa_{Z}^{2} \kappa_{B}^{2}}{0.1775 + 0.5809 \kappa_{B}^{2} + 0.2416 \kappa_{Z}^{2}}\,.
\end{equation}

In evaluating Eqn. \ref{ee4}, both the Higgs couplings with the bottom quark as well as with $Z$ bosons have been modified keeping coupling with other fermions unchanged. $\kappa_Z$ appears from the production cross-section, while $\kappa_B$ appears from the decay branching fraction, pertaining to the fact that 
Higgs production and decay modes are those predicted in the SM, the NP effects only enters via rescaling of the Higgs coupling strength. 
The denominator comes from the modification of the Higgs decay width, which is easily adjusted within the observed total Higgs decay width at 
LHC for moderate values of $\kappa$ parameters. We can see that for $\kappa_{Z}=\kappa_B=1$, the signal strength turns one.  
We further invoke the custodial symmetry requirement $\kappa_{W} = \kappa_{Z}$. This assumption reduces the number of free parameters in the model and is consistent with the LHC bounds reported in \cite{ATLAS:2022vkf, CMS:2022dwd}. However, this assumption can be relaxed, and the impact of doing so in our analysis is discussed in Appendix \ref{E}. The bounded region is shown in Figure \ref{fig:n1}. Additionally, superimposing the current correlated bound on the $\kappa_{V} - \kappa_{F}$ space from the combined Higgs measurement gives the status of the most accurate constraint from the current data. The uncertainties are less than 10\% from the SM prediction. 

\begin{figure}[htbp]
	\centering
	\includegraphics[trim=0 0 1.25cm 1.25cm, clip, width = 0.6 \textwidth]{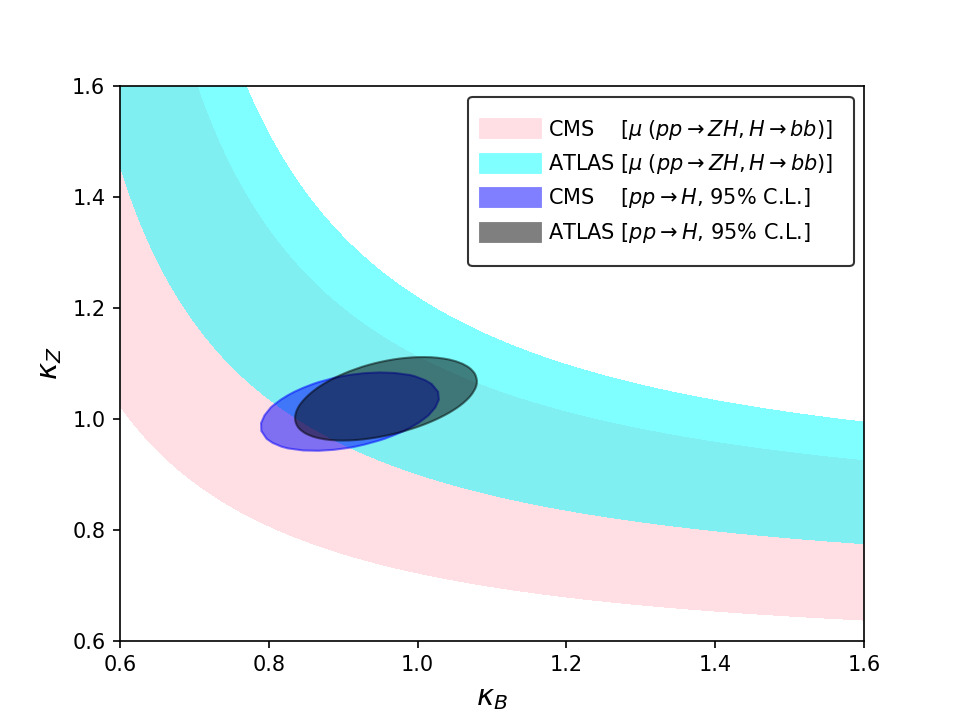}
	\hfill
	\caption{Constraints on coupling modifiers $\kappa_{Z}$ and $\kappa_{B}$ based on combined limit on $pp \xrightarrow{} Zh$, $h \xrightarrow{} b \Bar{b}$ production signal strength (bands) \cite{ATLAS:2020fcp,CMS:2018nsn} as well as the combined experimental best fit from all Higgs production and decay processes (ellipses) \cite{ATLAS:2022vkf,CMS:2022dwd} at ATLAS and CMS experiments. \label{fig:n1}}
\end{figure}


\subsection{Constraints on the effective coupling coefficients}

In context of the study of Higgs couplings in effective framework, several collaborations \cite{Ethier:2021bye,Ellis:2020unq,Brivio:2022hrb,Anisha:2021hgc,Buckley:2015lku} has attempted to provide global bounds based on Higgs, Top and Diboson data from Tevatron, LHC and LEP. Similar fits have been done including B-Physics data with LHC top and bottom quark studies \cite{Bissmann:2020mfi,Grunwald:2023nli}. Similarly, global fits have also been performed based on Electroweak Precision data as well as $\Delta_{\rm CKM}$ measurement \cite{ThomasArun:2023wbd}. Customarily, both ATLAS and CMS have done a number of studies on constraining the EFT operators across various channels using latest data. While ATLAS uses SMEFT, CMS uses HEL parametrization, we will consider bounds from ATLAS \cite{ATLAS:2020fcp,ATLAS:2021kog,ATLAS:2021vrm,ATLAS:2022xyx} only. Also, there have been numerous studies concerning SMEFT, projecting constraints for future collider experiments \cite{deBlas:2022ofj,Celada:2024mcf,Bissolotti:2023vdw}. We tabulate the constraints on the coefficients $c_{uW}$ and $c_{Hq}^{(3)}$ from current data in Table \ref{tab:03} and \ref{tab:03a}. The contraints on other dipole and Higgs-current operators are listed in Appendix \ref{D}.

\begin{table}[htbp]
	\centering
	\begin{tabular}{|c|c|c|c|}
		\hline
		Coefficient & Bound & C.L. & Flavor Scheme [Source] \\ \hline
		\multirow{6}{*}{$c_{uW}$} & [-0.375, 0.375] & 95\% & $U(3)_{q} \times U(3)_{u} \times U(3)_{d}$ \cite{Boughezal:2021tih} \\
		& [-0.290, 0.290] & 95\% & $U(2)_{q} \times U(2)_{u} \times U(2)_{d}$ \cite{daSilvaAlmeida:2019cbr} \\
		& [-0.084, 0.029] & 95\% & $U(1)_{Q} \times U(1)_{t}$ \cite{Ethier:2021bye} \\
		& [-0.012, 0.048] & 90\% & $U(3)_{q} \times U(3)_{u} \times U(3)_{d}$ \cite{Grunwald:2023nli} \\
		& [-0.079, 0.109] & 95\% & $U(1)_{Q} \times U(1)_{t}$ \cite{Buckley:2015lku} \\
		& [-0.120, 0.510] & 95\% & $U(1)_{Q} \times U(1)_{t}$ \cite{Ellis:2020unq} \\
		\hline
	\end{tabular}
	\caption{Constraints on dipole operator coefficient $c_{uW}$ from existing studies, see the references for details. $q$, $u$ and $d$ are the SM LH quark doublet, RH up and down-type quark singlets, respectivey. $Q$ and $t$ refers to the third generation LH quark doublet and RH Top quark singlet, respectively. \label{tab:03}}
\end{table}

\begin{table}[htbp]
	\centering
	\begin{tabular}{|c|c|c|c|}
		\hline
		Coefficient & Bound  & C.L. & Flavor Scheme [Source] \\ \hline
		\multirow{10}{*}{$c_{Hq}^{(3)}$} & [-0.432, 0.062] & 95\% & $U(2)_{q} \times U(2)_{u} \times U(3)_{d}$ \cite{Ethier:2021bye} \\
		& [-0.375, 0.344] & 95\% & $U(1)_{Q} \times U(1)_{t}$ \cite{Ethier:2021bye} \\
		& [-0.702, 0.104] & 68\% & $U(3)_{q} \times U(3)_{u} \times U(3)_{d}$ \cite{ThomasArun:2023wbd} \\
		& [-0.080, 0.100] & 95\% & $U(3)_{q} \times U(3)_{u} \times U(3)_{d}$ \cite{ATLAS:2021vrm} \\
		& [-0.018, 0.026] & 68\% & $U(3)_{q} \times U(3)_{u} \times U(3)_{d}$ \cite{ATLAS:2020fcp} \\
		& [-0.150, 0.520] & 95\% & $U(3)_{q} \times U(3)_{u} \times U(3)_{d}$ \cite{ATLAS:2021kog} \\
		& [-0.023, 0.053] & 90\% &  $U(3)_{q} \times U(3)_{u} \times U(3)_{d}$ \cite{Grunwald:2023nli} \\
		& [-0.017, 0.012] & 95\% & $U(2)_{q} \times U(2)_{u} \times U(3)_{d}$ \cite{Ellis:2020unq} \\
		& [-0.032, 0.048] & 95\% & $U(1)_{Q} \times U(1)_{t}$ \cite{Ellis:2020unq} \\
		& [-0.008, 0.016] & 95\% & $U(2)_{q} \times U(2)_{u} \times U(3)_{d}$ \cite{Anisha:2021hgc} \\
		\hline
	\end{tabular}
	\caption{Constraints on Higgs-current operator coefficient $c^{(3)}_{Hq}$ from existing studies, see the references for details. \label{tab:03a}}
\end{table}

A few comments and clarifications on the existing bounds: The bounds in Table \ref{tab:03} and \ref{tab:03a} are individual bounds for the operator coefficients, where all other operator coefficient values are set to zero. The bounds for some of the studies in Table \ref{tab:03} and \ref{tab:03a} are provided for $\left(c / \Lambda^{2}\right)$ or $\left(c v^{2} / \Lambda^{2}\right)$. The values provided on the tables are normalized to $\Lambda = 1$ TeV. The flavor schemes associated with different analyses are tabulated as well in the tables. The bounds significantly change as per flavor assumptions. In this analysis, we will assume a $U(2)_{q} \times U(2)_{u} \times U(2)_{d}$, stating that the third generation is completely decoupled from the first two generations. This assumption is guided by the fact that the collisions at LHC contributing to $Zh$ production are evidently light quarks (b quark collisions are suppressed by PDFs). The bounds are from different analyses at difference confidence levels. For our analysis in suceeding sections, we consider the EFT benchmarks: $c_{uW} = 0.01, 0.05$ and $c^{(3)}_{Hq} = \pm 0.01, \pm 0.05$. Since, only the flavor scheme of \cite{daSilvaAlmeida:2019cbr} match with our flavor scheme, we only adhere to their bounds for $c_{uW}$. For $c^{(3)}_{Hq}$, none of the flavor scheme matches with our choice and we need not adhere to any of them. Still numerically, for $c_{uW}$, the benchmark 0.01 is consistent with all the bounds in Table \ref{tab:03}, while the benchmark 0.05 is consistent with \cite{Boughezal:2021tih,daSilvaAlmeida:2019cbr,Buckley:2015lku,Ellis:2020unq} only. Similarly, for $c^{(3)}_{Hq}$, the benchmark 0.01 is consistent with all the bounds in Table \ref{tab:03a}, while $-0.01$ is not consistent with \cite{Anisha:2021hgc}. The benchmarks $0.05$ and $-0.05$ are more wishful, and is incompatible with \cite{ATLAS:2020fcp,Ellis:2020unq,Anisha:2021hgc} and \cite{ATLAS:2020fcp,Ellis:2020unq,Anisha:2021hgc,Grunwald:2023nli}, respectively, caveat to the flavor assumptions. The CP-violating gauge-Higgs operators listed in Table \ref{tab:1} are constrained by the fermion dipole moment observables, such as electron and neutron dipole moments, as they contribute at one loop to dipole operators. The current limits on the electron and neutron dipole moments are $1.1\times 10^{-29}$ and $3.0\times 10^{-26}$ (in units of $e$ cm), respectively \cite{Cirigliano:2019vfc}. These operators are also constrained by the measurements of the triple gauge-interactions of the form $WW\gamma$ and $WWZ$ at LEP and the CP asymmetry measured in the $B\to X_s \gamma$ decay. The CP asymmetry measured in  $B\to X_s \gamma$ decay is $A_{B\to X_s \gamma}=0.015(20)$. Bound on the imaginary part of dipole operator $c_{uW}$ from neutron EDM \cite{Kley:2021yhn} (assuming $U(2)_{q} \times U(2)_{u} \times U(2)_{d}$ flavor scheme) is shown below:
	\begin{equation}
		\begin{split}
			\mathfrak{Im}(c_{uW}) & < 0.000672 \\
		\end{split}
	\end{equation}
The bound on other dipole operators are listed in Appendix \ref{D}. The bounds are extremely stringent, hence difficult to probe at collider experiments. For our case, we restrict ourselves to the real dipole coefficients, hence evading these bounds. The bounds on light quark dipole operators from electron EDM appear at two loop involving a Yukawa interaction in the loop and hence are insignificant in comparison to other constraints \cite{Panico:2018hal}.


\section{Collider analysis: current and future runs} \label{s4}
\begin{figure}[htbp]
\centering
\includegraphics[trim=0 0 1.25cm 1.25cm, clip, width = 0.6 \textwidth]{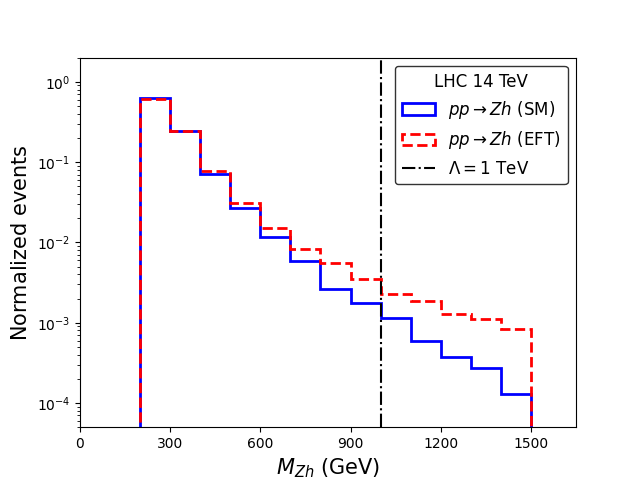}
\caption{Invariant mass of $Zh$ for the parton level processes in absence as well as presence of EFT operators at $\Lambda =$ 1 TeV.\label{figeval}. Here we consider the EFT benchmark point corresponding
to $\{\kappa_Z, c_{uW},c_{Hq}^{(3)}\} = \{1.00, 0.05, 0.05\}$.}
\end{figure}

In this section, we will discuss the event generation and analysis in the context of the current and future HL LHC data. 
We have simulated $pp \to Zh \to (l^{+} l^{-})(b\bar{b})$ process at the LHC 13 TeV and 14 TeV center of mass energies. 
The PDF set used is \texttt{NN23LO1} \cite{Ball:2012cx} and the renormalization and factorization scales are set at the 
default dynamical scale choice of MadGraph. The signal and background informations will be detailed in Section \ref{sbp}. 
Before proceeding for a collider analysis in presence of the EFT operators, it is important to check validity of the effective 
approximation for the collider setup. The validity condition is $\sqrt{\hat{s}} < \Lambda$, where $\sqrt{\hat{s}}$ is the partonic 
centre of mass energy. However, ensuring this criteria is difficult in hadron colliders like LHC, where collisions between partons 
happen at variable centre of mass energies. One way to deal with this is to construct the invariant mass of the final state particles, $M_{Zh}$ to guess the energy scale at which the partonic collisions take place. This is depicted in Figure \ref{figeval}, for the production process at a particular EFT benchmark point and we observe that almost all the events obey the effective validation criteria, when $\Lambda=1$ TeV. Now, the generated signal and background events are fed in \texttt{Pythia8} \cite{Sjostrand:2007gs} for parton showering. The showered events are fed in \texttt{Fastjet3} \cite{Cacciari:2011ma} for jet clustering using the \texttt{anti $k_{t}$} algorithm \cite{Cacciari:2008gp} with jet radius 0.5 and jet 
$p_{T} \geq$ 20 GeV. Finally, the events are processed through \texttt{Delphes3}, using the card \texttt{delphes\_card\_cms}, to take into account finite detector resolution effects, as well as 
particle identification, reconstruction and jet tagging. The processed events are analyzed using \texttt{MadAnalysis5} \cite{Conte:2012fm}.

\subsection{Cut based analysis \label{sbp}}

We consider the signal to be $pp \xrightarrow{} Z (l^+ l^-) H (b \Bar{b})$ i.e. a dilepton di-bjet signal. This signal is comparatively 
clean due to the presence of charged leptons with invariant mass of the dilepton system peaking around $Z$-mass. 
The possible SM backgrounds for this signal process are $pp \xrightarrow{} t \Bar{t}$, $pp \xrightarrow{} ZZ$, $gg \xrightarrow{} Zh$ and $pp \xrightarrow{} Z b \Bar{b}$.  
NLO corrections to both signal and background processes are taken into account by multiplying the production cross sections for each process by its NLO K factor, 
where $K = \sigma_{NLO}/\sigma_{LO}$. The LO and NLO cross sections are calculated using \texttt{MG5\_aMC@NLO} at LHC 13 TeV and the K factors for 
different processes are: $Zh$: 1.288 (for $pp \xrightarrow{} Zh$), 2.000 (for $gg \xrightarrow{} Zh$), $Zb \Bar{b}$: 1.230, $t \Bar{t}$: 1.338 and $ZZ$: 1.461 \cite{Cascioli:2014yka,Chen:2022rua,Campbell:2006wx}.

\begin{figure}[htbp]
	\centering
	\includegraphics[width=0.45 \textwidth]{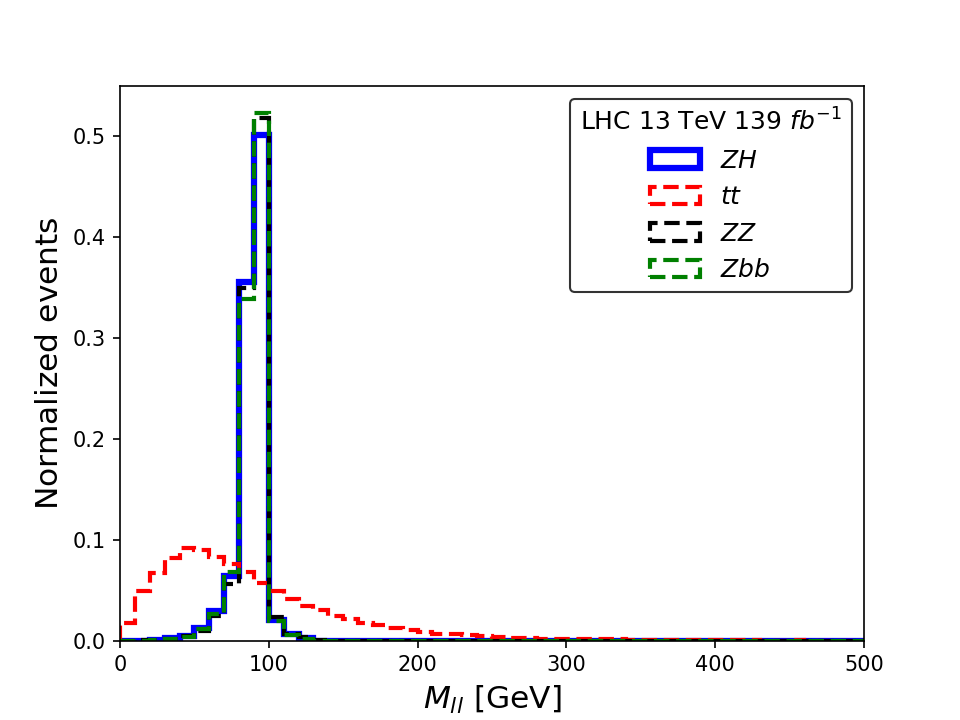}
	\includegraphics[width=0.45 \textwidth]{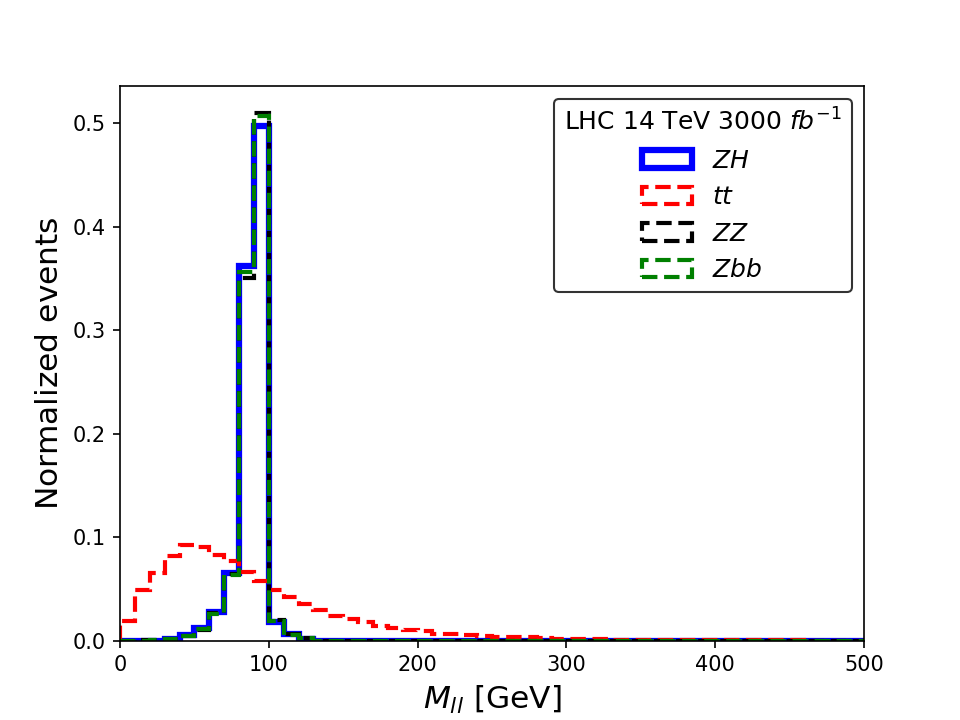} \\
	\includegraphics[width=0.45 \textwidth]{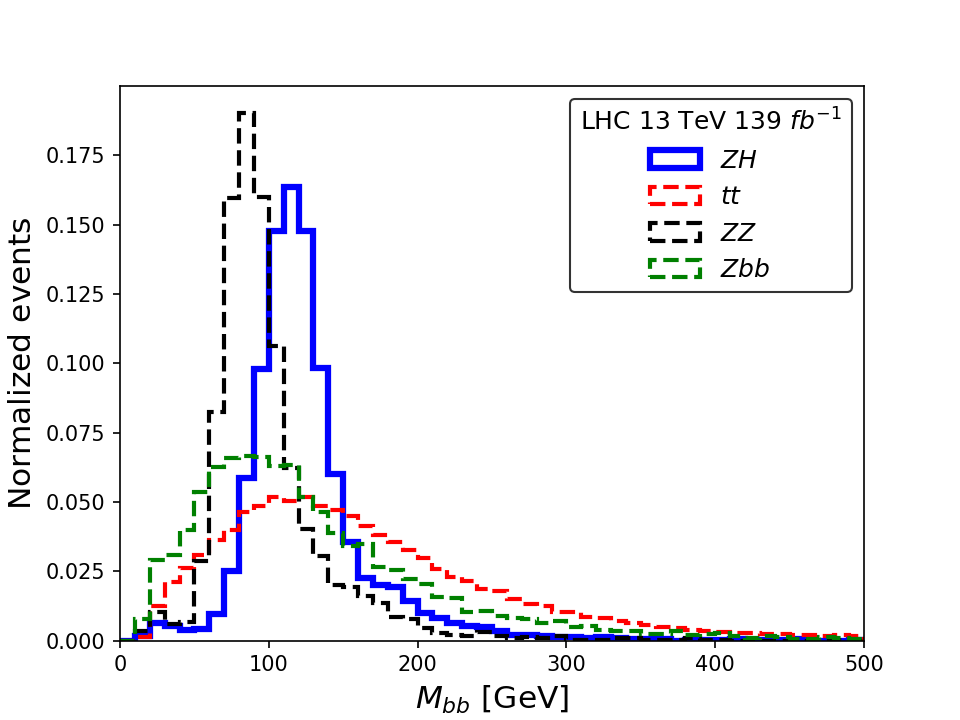}
	\includegraphics[width=0.45 \textwidth]{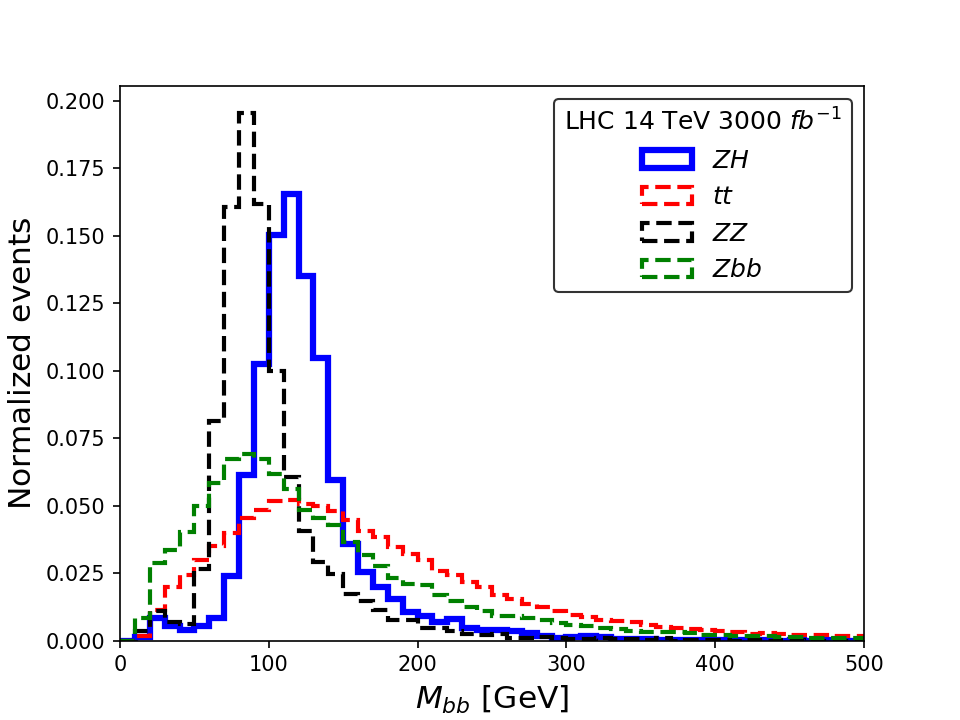} \\
	\includegraphics[width=0.45 \textwidth]{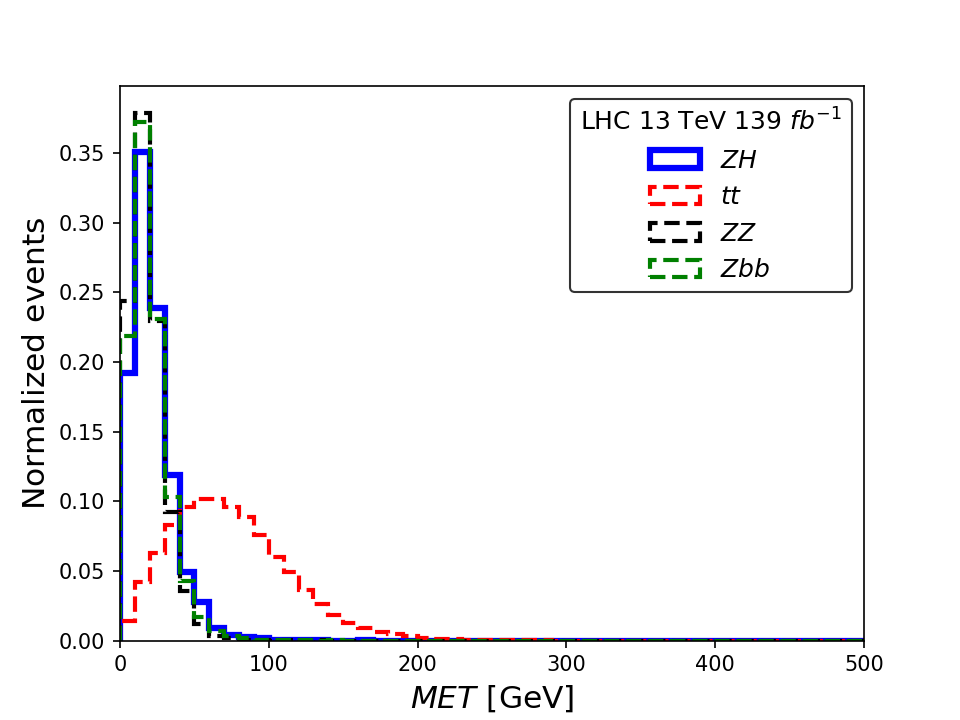}
	\includegraphics[width=0.45 \textwidth]{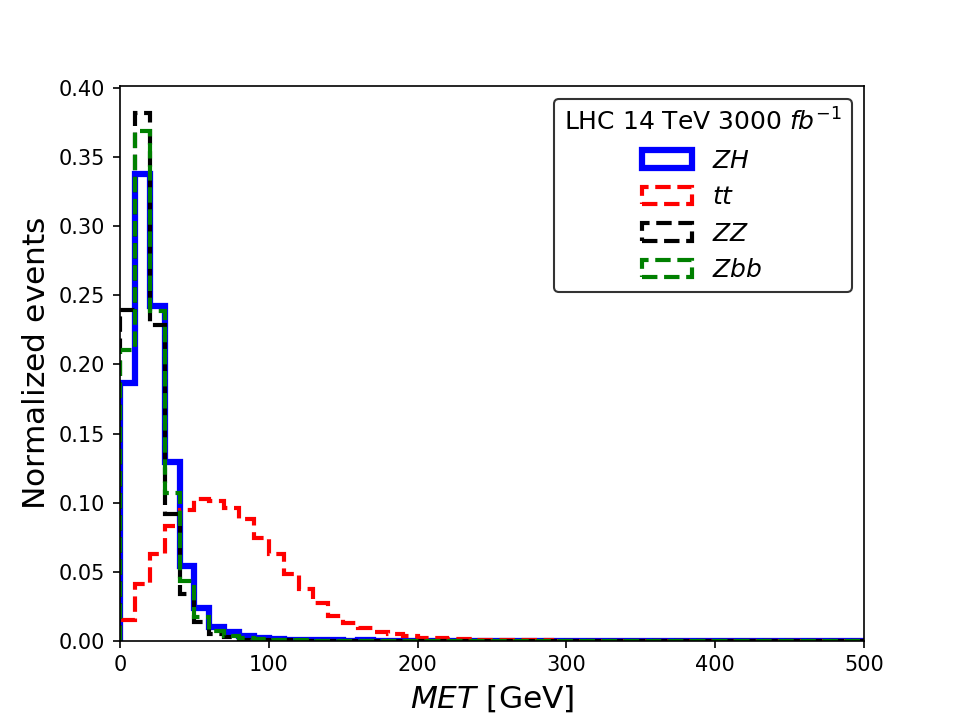} \\
	\caption{\textit{Top Left:} Invariant mass of dilepton at the LHC 13 TeV corresponding to the SM signal and backgrounds. \textit{Top Right:} Same for LHC 14 TeV. \textit{Middle Left:} Invariant mass of di-bjet  at the LHC 13 TeV. \textit{Bottom Right:} Same for LHC 14 TeV. \textit{Bottom Left:} Missing transverse energy at the LHC 13 TeV. \textit{Bottom Right:} Same for LHC 14 TeV. \label{fig:9}}
\end{figure}

We perform the analysis for two different setups viz. LHC 13 TeV 139 $fb^{-1}$ (equivalent to Run II) and LHC 14 TeV 3000 $fb^{-1}$ (future HL-LHC projection). 
We introduce some basic cuts common for all the analysis. These can be detailed as: $P_{T}^{j}$ $>$ 20.0 GeV, $P_{T}^{l}$ $>$ 10.0 GeV, -5.0 $<$ $\eta_{j}$ 
$<$ 5.0, -2.5 $<$ $\eta_{l}$ $<$ 2.5, $\Delta R_{jj}$ $>$ 0.4, $\Delta R_{lj}$ $>$ 0.4, $\Delta R_{ll}$ $>$ 0.4, where, subscripts $j$ and $l$ refer to jets and leptons 
respectively, and the kinematic variables have their usual definitions as in collider literature \cite{Schwartz:2017hep}. Apart from these, to ensure exclusive 
$l^{+}l^{-}b \Bar{b}$ signal we select events with $N_{b} = 2$ and $N_{l} = 2$. The relevant distributions of kinematic variables and observables are plotted 
in Figure \ref{fig:9} for 13 TeV and 14 TeV LHC. Choosing $N_j < 5$ for $Zh$ considerably reduces the pure hadronic backgrounds at LHC. We further choose a wide $M_{bb}$ cut around the Higgs mass to make sure that we retain most of the signal while eliminating the backgrounds. The presence of dileptons enables 
us to use some additional cuts. Owing to the better detectability of leptons, a narrow cut on $M_{ll}$ around the $Z$ mass significantly reduces the
 $t\bar{t}$ background. Finally due to the absence of invisible particles in the final state (a part from neutrinos possibly coming $b$-decay) rejecting 
 events with large missing transverse energy, MET (MET $> 30$ GeV) removes backgrounds with neutrinos in final state like leptonic $t \Bar{t}$ mode. The entire cut flow for SM 
 Zh process (in absence of EFT contributions) is tabulated in Table \ref{tab4}.

\begin{table}[htbp]
	\centering
	\resizebox{\textwidth}{!}{
	\begin{tabular}{|l|c|c|c|c|c|c|c|c|}
		\hline
		{} &  \multicolumn{4}{c|}{LHC 13 TeV 139 $fb^{-1}$} & \multicolumn{4}{c|}{LHC 14 TeV 3000 $fb^{-1}$} \\ \hline
		 Cuts & S & B & S/B & z & S & B & S/B & z \\
		 \hline
		Basic Cuts & 569 & 1016890 & 0.0006 & 0.56 & 13327 & 25581580 & 0.0005 & 2.63 \\
		$N_{j}<$ 5 & 309 & 369549 & 0.0008 & 0.51 & 7054 & 9079825 & 0.0008 & 2.34 \\
		$M_{bb} \in$ (110.0, 140.0) GeV& 126 & 56059 & 0.0022 & 0.53 & 2858 & 1382411 & 0.0021 & 2.43 \\
		$M_{ll} \in$ (80.0, 100.0) GeV& 107 & 11674 & 0.0092 & 0.99 & 2456 & 285269 & 0.0086 & 4.59 \\
		MET $<$ 30 GeV& 91 & 5472 & 0.0166 & 1.23 & 2026 & 129095 & 0.0156 & 5.62 \\
		\hline
	\end{tabular}
	}
		\caption{Cut flow and signal significance (z) for the SM $Z(l^{+}l^{-})h(b \Bar{b})$ associated production signal with the SM background processes at 
		LHC with $\sqrt{s}=$13 TeV and luminosity of 139 $fb^{-1}$ and at LHC with $\sqrt{s}=$ 14 TeV and integrated luminosity of 3000 $fb^{-1}$. Apart, we 
		use the following notations, S: Signal events, B: Background events, S/B: Signal to background ratio. \label{tab4}}
\end{table}

\begin{table}[htbp]
	\centering
	\begin{tabular}{|c|c|c|c|c|c|c|}
		\hline
		BPs $\left(\kappa_{Z}, c_{uW}, c_{Hq}^{(3)}\right)$ & S & B & S/B & z & $\epsilon_{s}$ & $\epsilon_{b}$ \\ \hline
		SM (1.00, 0.00, 0.00) & 2026 & \multirow{4}{*}{129095} & 0.0156 & 5.62 & $2.09 \times 10^{-2}$ & \multirow{4}{*}{$8.72 \times 10^{-4}$} \\
		BP1 (1.00, 0.05, 0.00) & 2142 &  & 0.0166 & 5.94 & $2.19 \times 10^{-2}$ &  \\
		BP2 (1.00, 0.00, 0.05) & 2238 & & 0.0173 & 6.21 & $2.10 \times 10^{-2}$ &  \\
		BP3 (1.00, 0.00, -0.05) & 1869 &  & 0.0145 & 5.19 & $2.09 \times 10^{-2}$ &  \\ \hline
	\end{tabular}
	\caption{Signal significance of the $Z(l^{+}l^{-})h(b \Bar{b})$ production process (at LHC with $\sqrt{s}=$14 TeV and luminosity 3000 $fb^{-1}$) after the final selection cuts listed in Table \ref{tab4} for different benchmark points (BPs) in presence of effective operators, along with signal ($\epsilon_{s}$) and background 
	($\epsilon_{b}$) efficiencies. \label{tab5}}
\end{table}
For signal significance, we use the Asimov definition \cite{Cowan:2010js}:
\begin{equation}
	z = \sqrt{2 \left((S + B)\ln\left(1 + \frac{S}{B} \right) - S \right)}\,,
\end{equation}
where $S$ is the number of signal events and $B$ is the number of background events. The Asimov significance is derived from the likelihood ratio, which takes into account the full statistical model, including the Poisson nature of signal and background events. Hence it is more statistically rigorous, not only for small event counts or comparable signal-background scenarios, but also in the general case. On Taylor expansion in the limit $S \ll B$, the Asimov significance approximates to $S/\sqrt{B}$, which is the more common definition. The number of signal and background events can further be parametrized in terms of the coupling modifiers and the effective operator coefficients as shown in Eqn. \ref{e42} .
\begin{equation} \label{e42}
	\begin{split}
		S &= \mathcal{L} \times \sigma_{SM + EFT}^{S} (\kappa_{V}, c_{i}) \times BR^{S} (\kappa_{B}, c_{i}) \times \epsilon_{S}\,, \\
		B &= \mathcal{L} \times \sigma_{SM}^{B} \times BR^{B} \times \epsilon_{B}\,; \\
	\end{split}
\end{equation}
where, $\mathcal{L}$ is the integrated luminosity, $\epsilon_{S}$ and $\epsilon_{B}$ are the signal and background cut efficiencies respectively. The EFT effects on $Z$ branching to leptons is parametrized in Appendix \ref{A}. From Table \ref{tab4}, it is clear that the signal significance improves significantly as we move to the high luminosity frontier.

The second part of the analysis features inclusion of EFT operators. In presence of EFT operators, there is a significant change in the signal significance. This has been illustrated in Table \ref{tab5} in presence of SMEFT operators at some benchmark values. The relevant kinematic variables and observables for the different benchmark points are shown in Figure \ref{fig:9a} of Appendix \ref{B}. The deviation from the SM signal is minuscule. The benchmarks are chosen at some standard values to gauge the effect on the overall parameter space.


\subsection{BDT based analysis}

With the advent of modern machine learning tools, multivariate analysis has become a norm in collider analysis \cite{Coadou:2022nsh,Cornell:2021gut}. 
Among the machine learning techniques, boosted decisions trees (BDT) based algorithms are extremely popular due to their stability as well as accuracy 
even for smaller datasets. With an aim to improved the signal background separation with more accuracy, we perform an analysis using the 
\texttt{XGBoost} classifier \cite{Chen_2016}. The core setup of the analysis is kept similar to the cut based counterpart. Since, we observed higher significance at 
14 TeV, 3000 $fb^{-1}$ LHC, we will perform the BDT based analysis exclusively for the HL-LHC run. The training sets for signal ($pp \rightarrow Z(l^+ l^-) H(b \Bar{b})$) 
and backgrounds ($pp \rightarrow t \Bar{t}$, $pp \rightarrow ZZ$, $gg \rightarrow Zh$ and $pp \rightarrow Z b \Bar{b}$) are prepared and weighted according 
to its number strength at 3000 $fb^{-1}$. This weighted data is subjected to the same basic cuts as used in the cut based analysis. The feature selection is done 
in similar line with the $Vh$, $h \rightarrow b \Bar{b}$ analysis at ATLAS \cite{ATLAS:2020fcp}. The features are detailed below:
\begin{itemize}
	\item Transverse momentum of the leading lepton: $p_{T}^{l}$
	\item Transverse momentum of the leading b-jet: $p_{T}^{b}$
	\item Number of jets (inclusive of b-jets): $N_{j}$
	\item Missing transverse energy: MET
	\item Transverse momentum of the dilepton system: $p_{T}^{ll}$
	\item Transverse momentum of the di-b-jet system: $p_{T}^{bb}$
	\item Invariant mass of the dilepton system: $M_{ll}$
	\item Invariant mass of the di-b-jet system: $M_{bb}$
	\item Separation in $(\eta, \phi)$ space between the leptons: $\Delta R^{ll}$
	\item Separation in $(\eta, \phi)$ space between the b-jets: $\Delta R^{bb}$
	\item Difference in $\eta$ between the dilepton and di-b-jet system: $\Delta \eta^{ll, bb}$
	\item Difference in $\phi$ between the dilepton and di-b-jet system: $\Delta \phi^{ll, bb}$ 	 
\end{itemize} 
The detailed correlation heatmap between the features for signal and backgrounds is shown in Appendix \ref{C}. To prevent features with large magnitudes (like $p_{T}$, MET, etc) to dominate over small valued features (like $\Delta \eta$, $\Delta R$, etc), all the features are scaled about their mean value ($\Bar{x}$) by their standard deviation ($SD$) i.e. $(x-\Bar{x})/SD$. This ensures all features are more or less at equal footing at the beginning of the training process. The signal and background events are divided in 1:1 ratio for training and testing each process. For training, the overall signal weight is equated with the overall background weight ensuring equal number of signal and background events entering the training process. The BDT model is constructed using \texttt{XGBClassifier} and the hyperparamter tuning is done using \texttt{GridSearchCV} \cite{scikit-learn}. The details of the optimal hyperparameters is listed in Table \ref{hp} in Appendix \ref{C}. The model is 3 fold cross validated using \texttt{StratifiedKFold} \cite{scikit-learn}. Since, the distributions for the SM plus EFT benchmarks vary very little from the SM distributions, we train only the SM signal and backgrounds and test on the SM as well as on SM plus EFT benchmarks.

\begin{figure}[htbp]
	\includegraphics[trim=0 0 1.25cm 0.5cm, clip, width = 0.475 \textwidth]{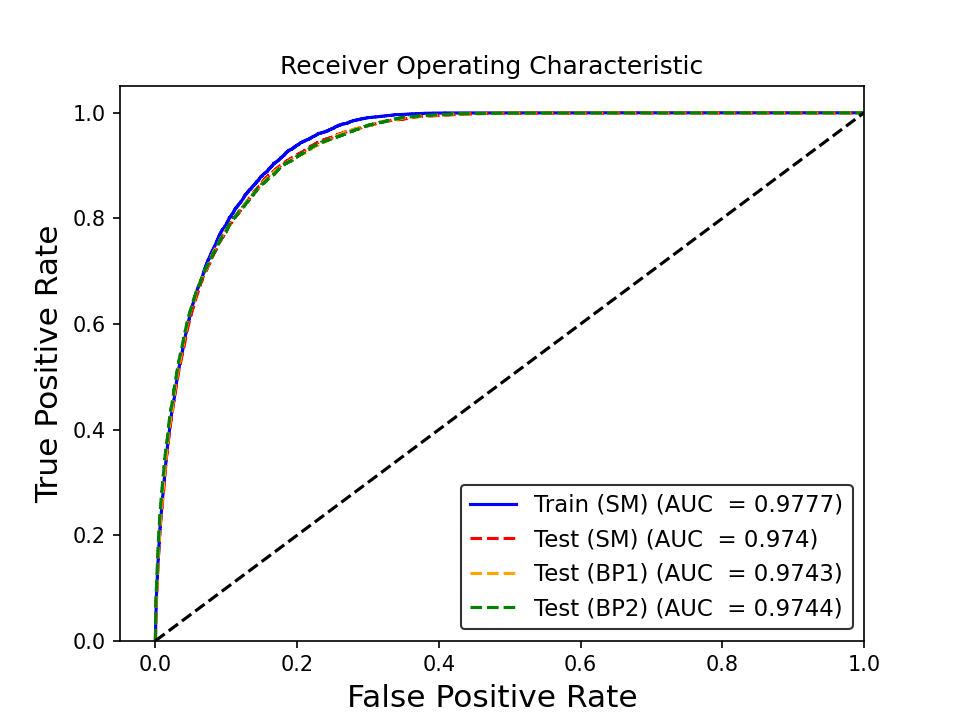}
	\includegraphics[trim=0 0 1.25cm 0.5cm, clip, width = 0.475 \textwidth]{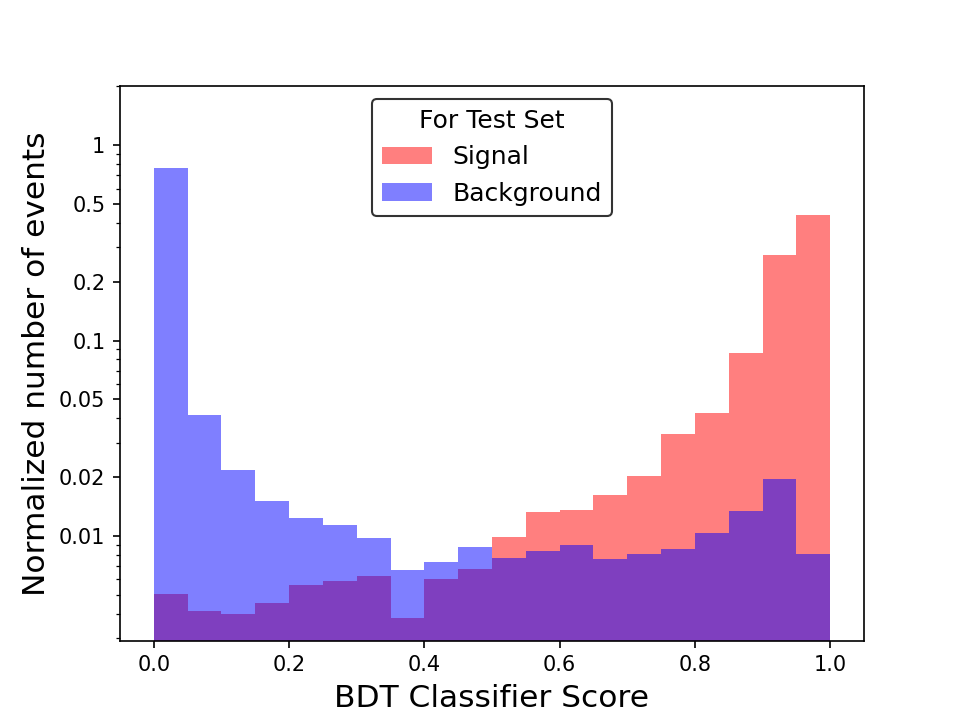}
	\includegraphics[trim=0 0 0cm 0cm, clip, width = 0.475 \textwidth]{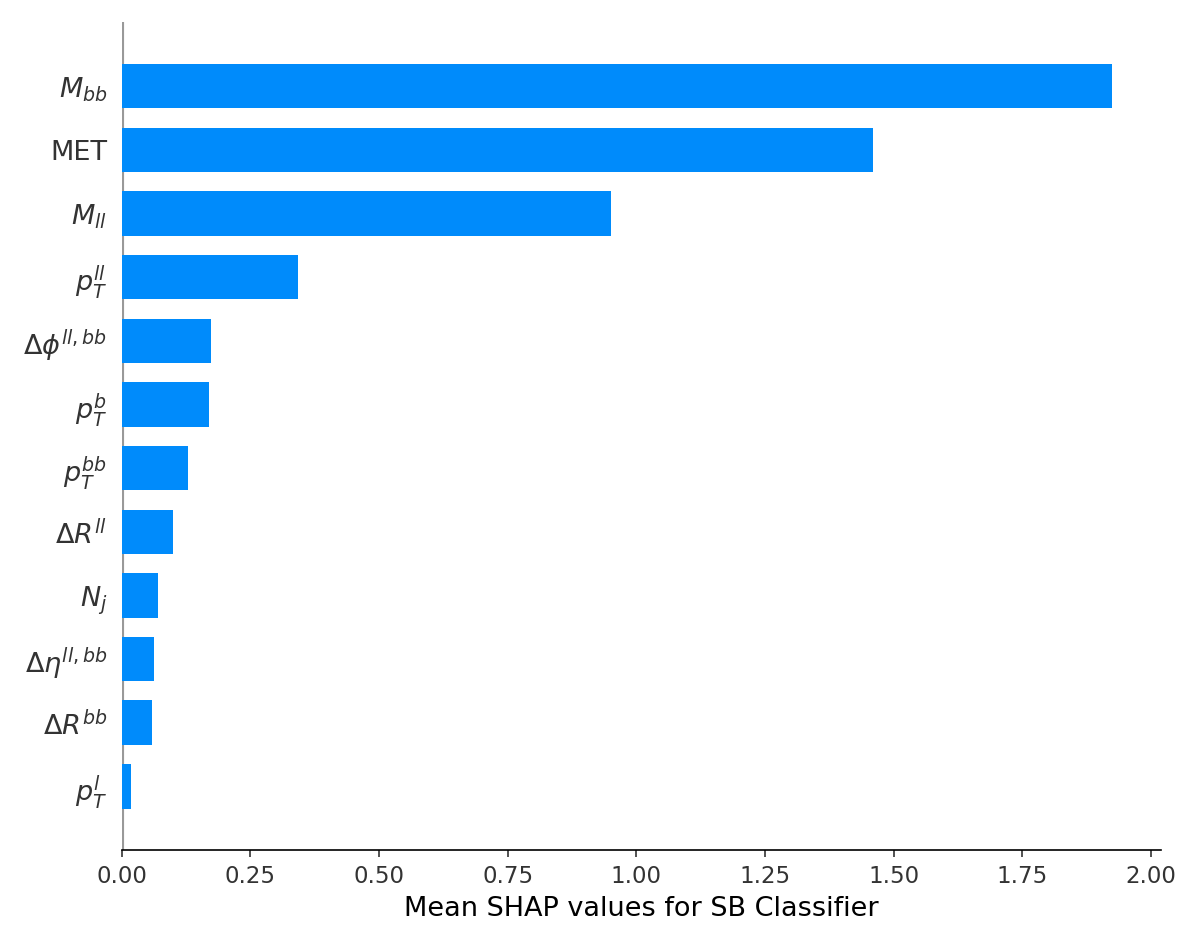}
	\hspace{0.5cm}
	\includegraphics[trim=0 0 1.25cm 1.25cm, clip, width = 0.475 \textwidth]{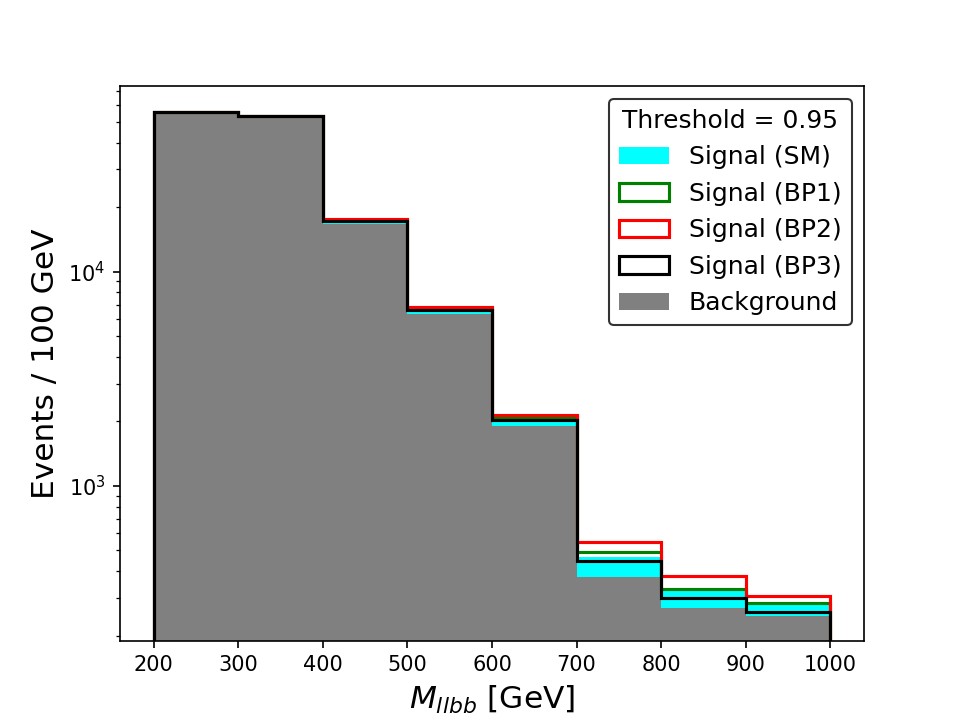}
	\caption{\textit{Top Left:} The Receiver Operating Characteristic (ROC) curve for the training and testing (SM as well as BPs) data. \textit{Top Right:} The distribution of the BDT classifier score for the SM signal and backgrounds. \textit{Bottom Left:} The mean absolute SHAP values for the features used for training the BDT model. \textit{Bottom Right:} $M_{llbb}$ distribution of signal and background processes post threshold selection of 0.95. \label{fig:12}}
\end{figure}

The highlights of the BDT analysis are shown in Figure \ref{fig:12}. As a metric for evaluation of our model, we use the ROC AUC. The AUC value for the training set is 0.9777. For the test sets, the AUC obtained are 0.9740, 0.9743 and 0.9744 respectively for SM, BP1 and BP2. This indicate that there was very little overfitting and the BPs fared very well on the model trained with SM inputs. To enhance the signal significance, we have set the threshold classifier score at 0.95. This means we will only keep events with a BDT classifier score greater than or equal to 0.95. By doing this, we maintain a roughly similar background efficiency as in the cut-based analysis, making it easier to compare signal efficiencies between the two methods. The feature importance of a machine learning model can be realized using SHAP (SHapley Additive exPlanations) \cite{lundberg2017unified} values for the features. SHAP values show how each feature affects each final prediction, the significance of each feature compared to others, and the model's reliance on the interaction between features. The SHAP values for the model are plotted in Figure \ref{fig:12} using \texttt{shap} module. It is distinctly observed that $M_{bb}$ turns out to be the best discriminating feature. MET, $M_{ll}$ and $p_{T}^{ll}$ turns out to be the other important features which was also taken into account in the cut based analysis. Corresponding to the threshold value of $0.95$, the significance and cut efficiencies of the signal and background processes for the BDT analysis are detailed in Table \ref{tab5b}. It is observed that for similar background numbers, there is a significant boost in the signal numbers for both SM and other EFT benchmark points, compared to the cut based analysis.

\begin{table}[htbp]
	\centering
	\begin{tabular}{|c|c|c|c|c|c|c|}
		\hline
		BPs ($\kappa_{Z}$, $c_{uW}$, $c_{Hq}^{(3)}$) & S & B & S/B & z & $\epsilon_{s}$ & $\epsilon_{b}$ \\ \hline
		SM (1.00, 0.00, 0.00) & 4048 & \multirow{4}{*}{129397} & 0.0313 & 11.20 & $4.17 \times 10^{-2}$ & \multirow{5}{*}{$8.74 \times 10^{-4}$} \\
		BP1 (1.00, 0.05, 0.00) & 4217 & & 0.0326 & 11.66 & $4.33 \times 10^{-2}$  &  \\
		BP2 (1.00, 0.00, 0.05) & 4762 & & 0.0368 & 13.16 & $4.47 \times 10^{-2}$ &    \\
		BP3 (1.00, 0.00, -0.05) & 3484 & & 0.0269 & 9.64 & $3.89 \times 10^{-2}$ &    \\
		\hline
	\end{tabular}
	\caption{Signal significance of the $Z(l^{+}l^{-})h(b \Bar{b})$ production process (at LHC 14 TeV 3000 $fb^{-1}$) from BDT based analysis (with threshold: 0.95) at different benchmark points (BPs) in presence of effective operators, along with signal ($\epsilon_{s}$) and background ($\epsilon_{b}$) efficiencies. \label{tab5b}}
\end{table}


\subsection{Two parameter $\chi^{2}$ analysis} \label{ss1}

The $\chi^{2}$ test \cite{ParticleDataGroup:2022pth} serves as a statistical measure to evaluate the degree of conformity between empirical observations 
and theoretical expectations within the framework of a contingency table. In statistical literature, there exist a number of definitions for the $\chi^{2}$ function. 
These definitions are more or less equivalent and can be used interchangeably. For our case, we shall use Pearson's definition where the $\chi^{2}$ is defined as 
follows:
\begin{equation}
	\chi^{2} = \sum_{j}^{bins} \left( \frac{\mathfrak{O}^{j} (\kappa_{i}, c_{i}) - \mathfrak{O}^{j}_{SM}}{\Delta N^{j}(\kappa_{i}, c_{i})} \right)^{2} = \sum_{j}^{bins} \left( \frac{\Delta \mathfrak{O}^{j} (\kappa_{i}, c_{i})}{\Delta N^{j} (\kappa_{i}, c_{i})} \right)^{2}\,.
\end{equation}
Here, $\mathfrak{O}$ is a differential distribution. For our analysis, $\mathfrak{O} = \left(d \sigma/d M_{llbb}\right)$, which is the invariant mass of all the visible final state particles following relevant cut/BDT analysis. We define $\Delta \mathfrak{O}^{j} (\kappa_{i}, c_{i}) = \mathfrak{O}^{j} (\kappa_{i}, c_{i}) - \mathfrak{O}^{j}_{SM}$, where $ \mathfrak{O}^{j} (\kappa_{i}, c_{i})$ corresponds to events in the $j^{th}$ bin of the $M_{llbb}$ distribution, in presence of effective operators and coupling modifiers and $\mathfrak{O}^{j}_{SM}$ corresponds to signal events in the same bin under the SM process. $\Delta N^{j} (\kappa_{i}, c_{i}) = \sqrt{\mathfrak{O}^{j} (\kappa_{i}, c_{i})+\mathfrak{O}^{j}_{SM}}$ is the statistical uncertainty in the number of events for that bin.
 
\begin{figure}[htbp]
	\centering
	\includegraphics[width = 0.475 \textwidth]{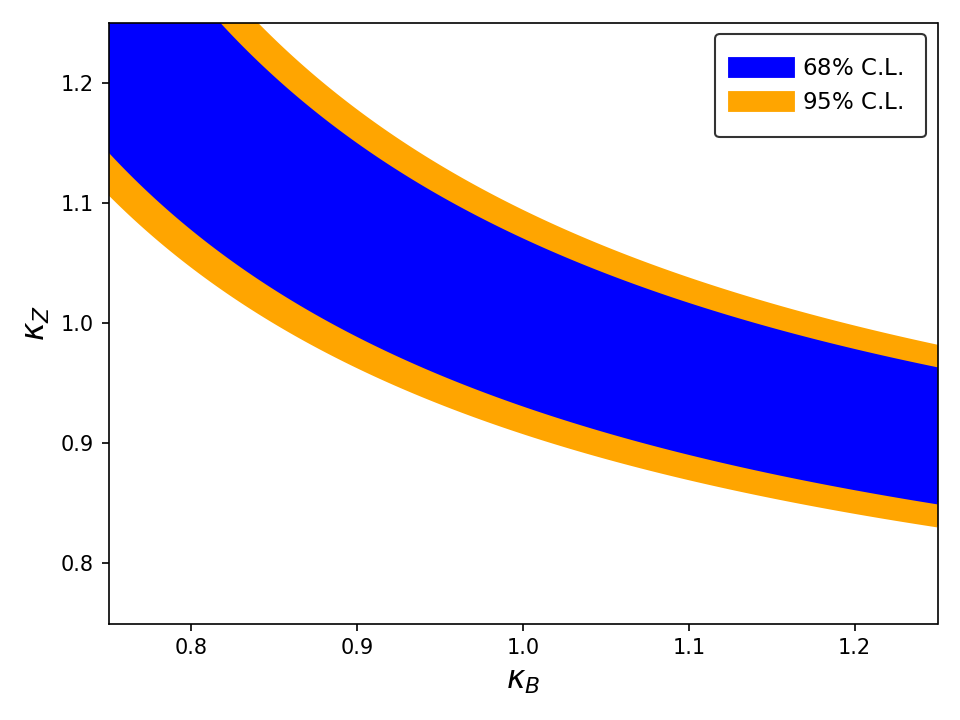}
	\includegraphics[width = 0.475 \textwidth]{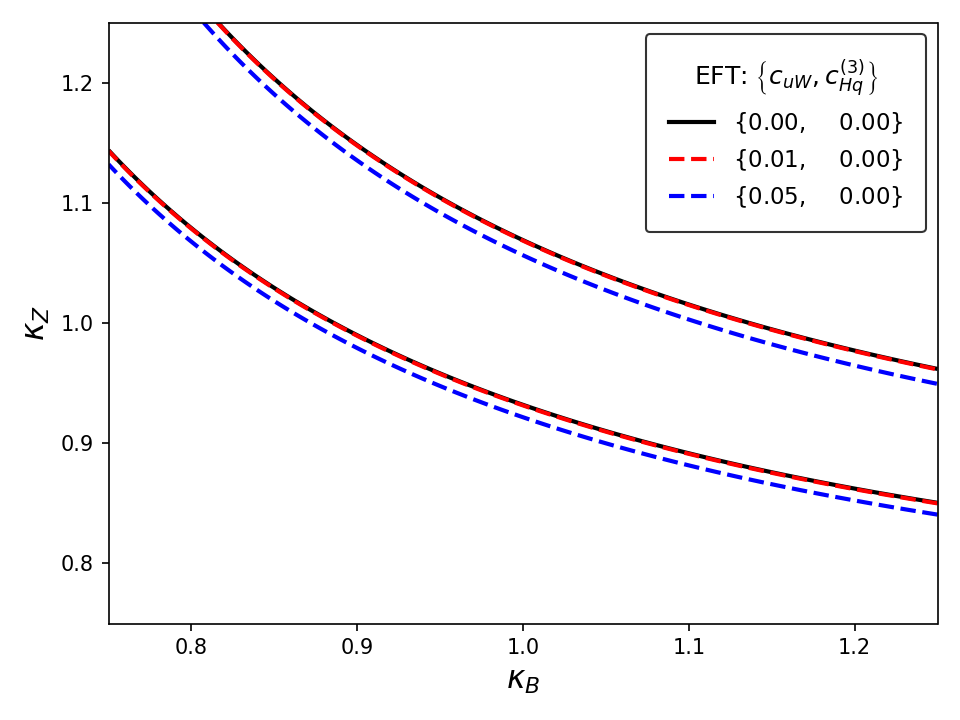}
	\includegraphics[width = 0.475 \textwidth]{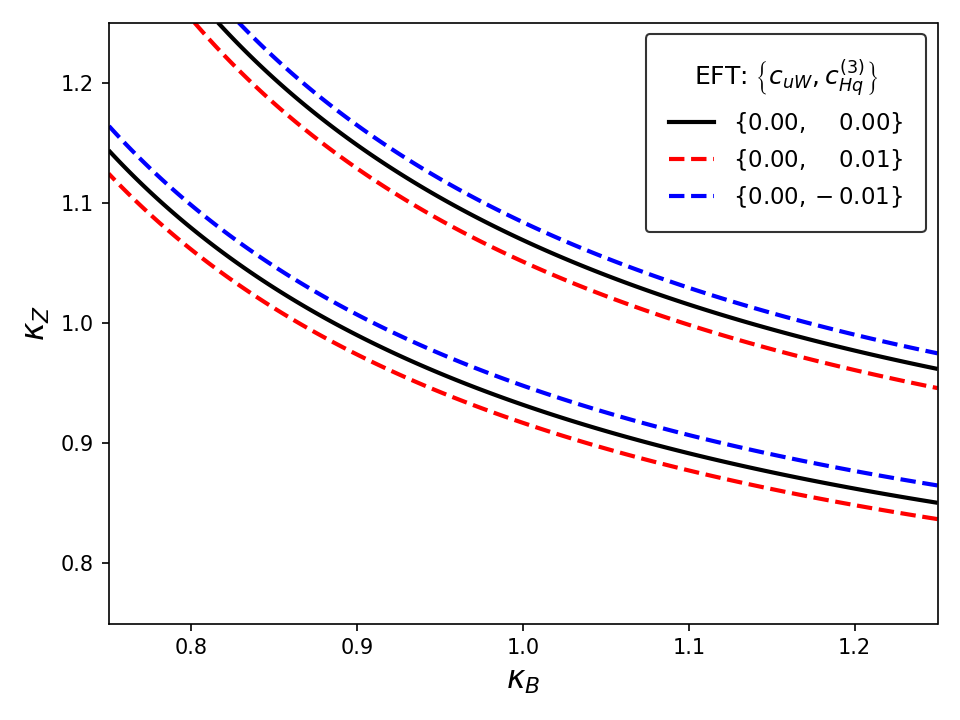}
	\includegraphics[width = 0.475 \textwidth]{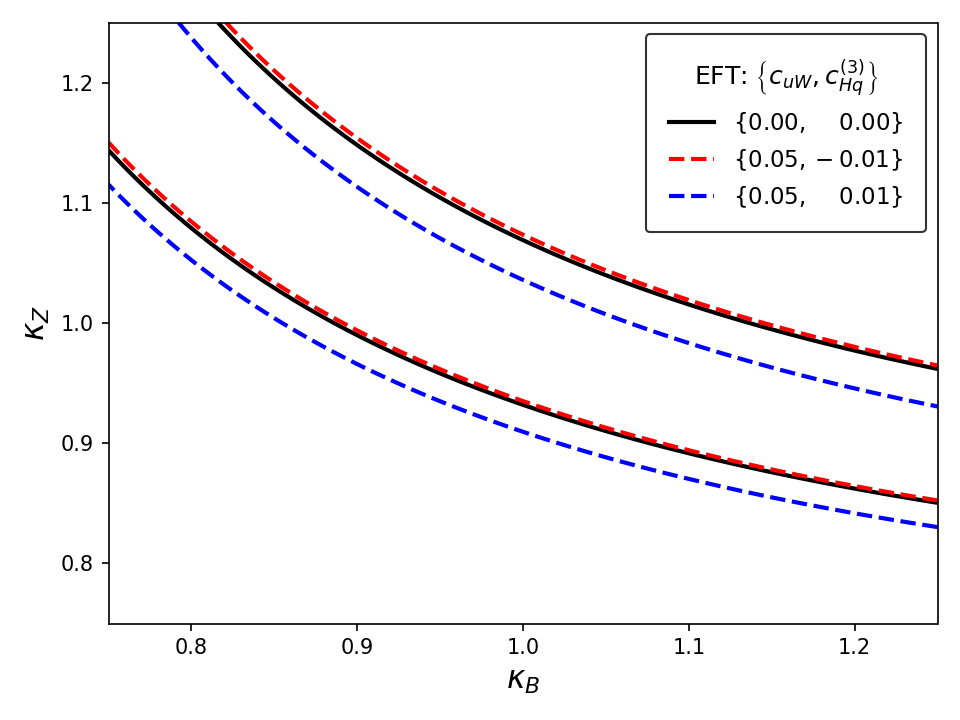}
	\caption{\textit{Top Left:} $\chi^{2}$ plots showing 68\% C.L. and 95\% bounds in the $\kappa_{Z}-\kappa_{B}$ space at LHC 14 TeV 3000 $fb^{-1}$. \textit{Top Right, Bottom Left} and \textit{Bottom Right:} Shift in 68\% C.L. bounds for different values of effective coefficients. \label{fig:14}}
\end{figure}

 Considering the better performance of the BDT analysis over the cut based analysis, we use the signal and background efficiencies based on the BDT analysis. The $\chi^{2}$ analysis is performed corresponding to the binned $M_{llbb}$ distribution. The dataset post the threshold cut is divided into 8 bins from 200-1000 GeV in intervals of 100 GeV. The events with higher values of $M_{llbb}$ are removed in order to adhere to the EFT limit. The BDT analysis is repeated for several benchmarks on the parameter space and the distributions corresponding to some of the benchmarks are shown in Appendix \ref{B}. For a Poisson distributed binned data, the number of degrees of freedom is given by $D = N - M$, where $N$ is the number of bins and $M$ is the number of fitted parameters. In this case, $D = 8 - 2 = 6$. Usually the constraints on the parameter space of physical parameters are reported in terms of confidence levels (C.L.). The confidence levels can be interpreted from the $\chi^{2}$ values. The 68\% and 95\% C.L. for $D = 6$ corresponds to $\chi^{2} \approx 7.01$ and $\chi^{2} \approx 12.59$ respectively. The $\chi^{2}$ for different parameter benchmarks in $\kappa_{Z}$ - $\kappa_{B}$ plane are shown in Figure \ref{fig:14}.
 
Fig. \ref{fig:14} shows that the 68\% C.L. bounds are moderately sensitive to the parameter $c_{Hq}^{(3)}$ corresponding to the Higgs-current interaction. This is because $\frac{\sigma}{\sigma_{SM}}$ changes significantly under the variation of  $c_{Hq}^{(3)}$ in the $Vh$ channel (see Figure \ref{fig:7} left). However, it requires higher values of $c_{uW}$ to show similar effect in the $\chi^{2}$ plots. This is mostly due to the fact that the dipole operators contribute at $(1/\Lambda^4)$ order. Further, it should also be noted that the EFT effects can negate each other as evident from the benchmark $\{c_{uW}, c^{(3)}_{Hq}\} = \{0.05, -0.01\}$ (see Figure \ref{fig:14} bottom right) where the positive contribution on the cross section from $c_{uW}$ is cancelled by the negative contribution from $c^{(3)}_{Hq}$ showing negligible deviation from the SM contribution even for non-zero EFT contributions.

\begin{figure}[htbp]
	\centering
	\includegraphics[width = 0.475 \textwidth]{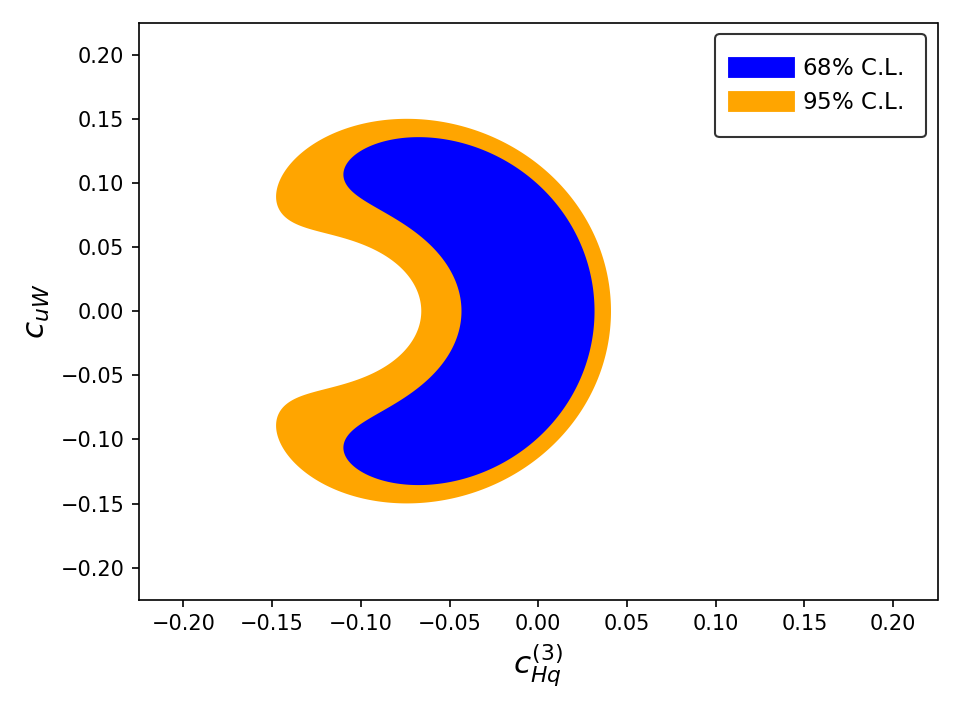}
	\includegraphics[width = 0.475 \textwidth]{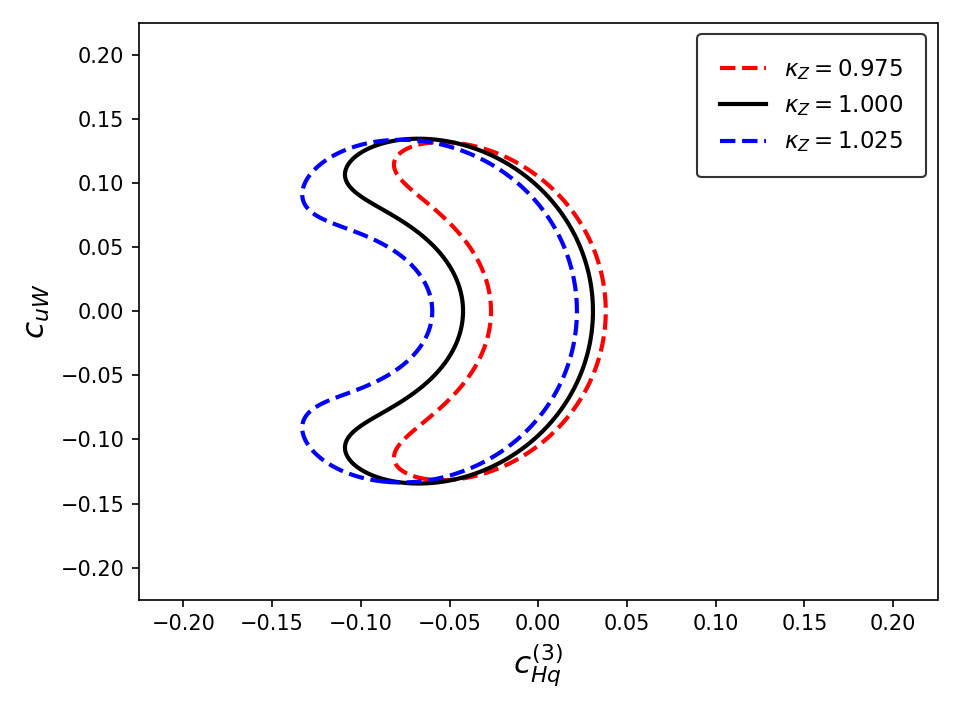} \\
	\includegraphics[width = 0.475 \textwidth]{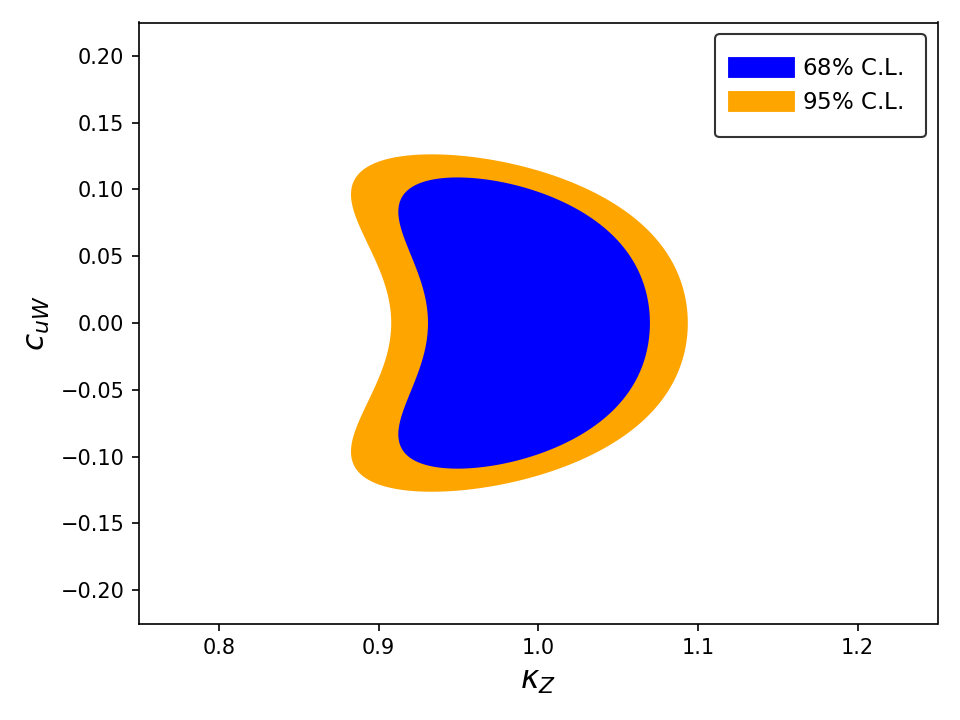}
	\includegraphics[width = 0.475 \textwidth]{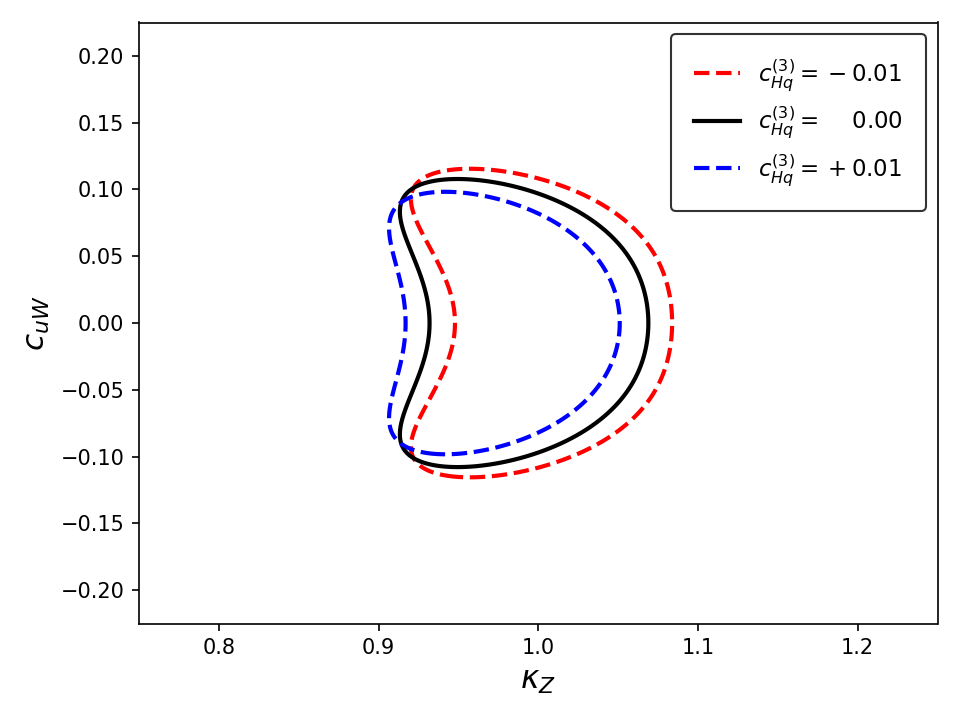} \\
	\includegraphics[width = 0.475 \textwidth]{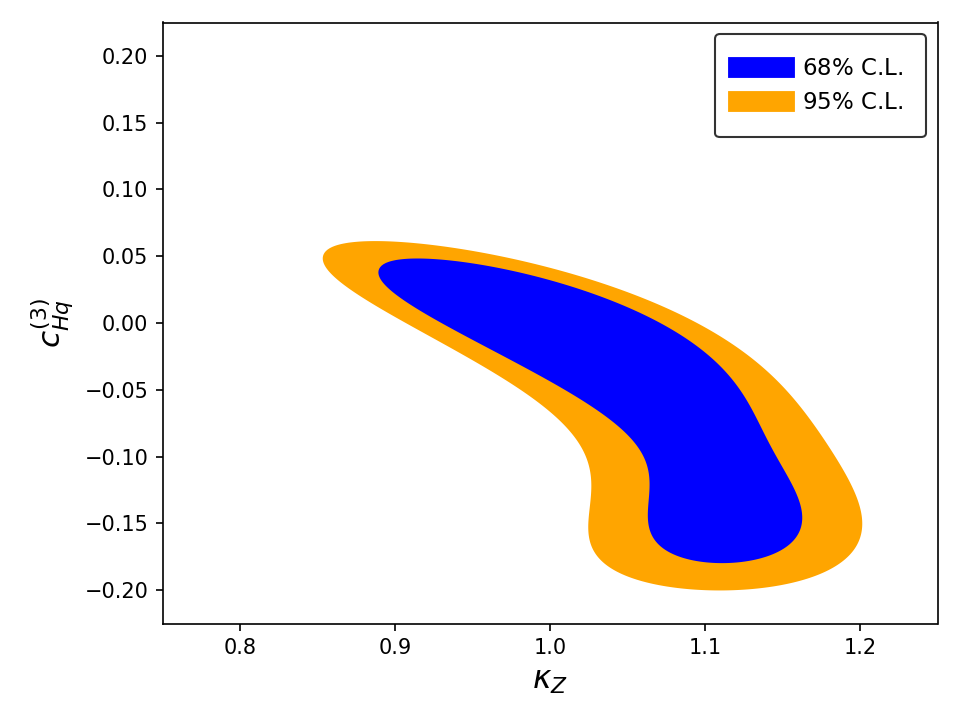}
	\includegraphics[width = 0.475 \textwidth]{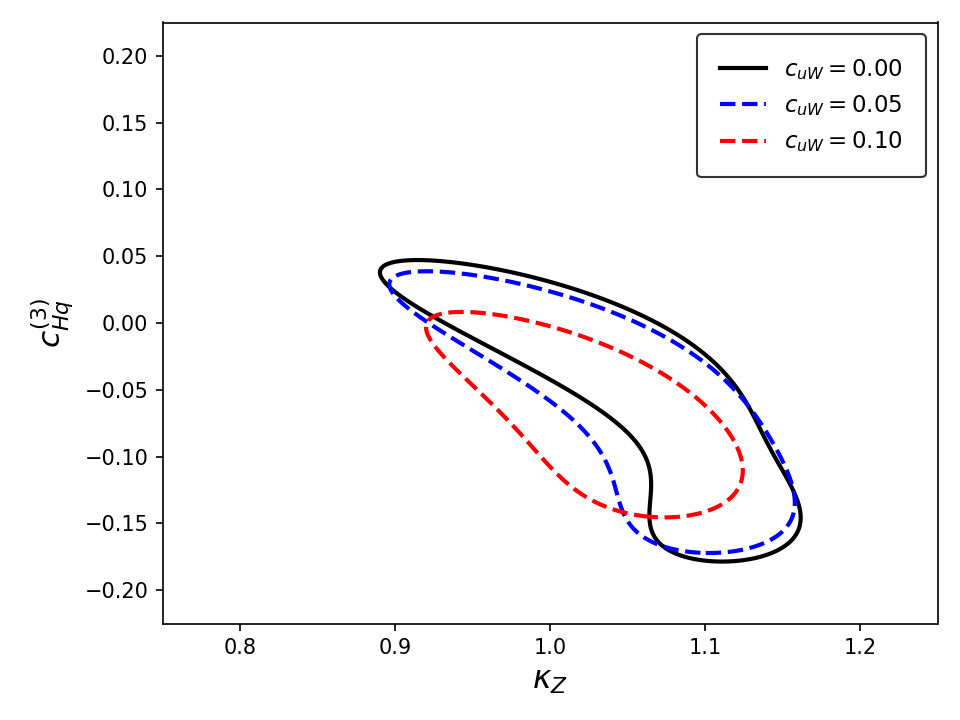} \\
	\caption{\textit{Top Left, Middle Left} and \textit{Bottom Left:} 68\% and 95\% C.L. bounds on effective coefficients at the 14 TeV 3000 $fb^{-1}$ LHC. \textit{Top Right, Middle Right} and \textit{Bottom Right:} Shift in 68\% C.L. in presence of different modifiers or EFT coefficients. \label{b2}}
\end{figure}

The 68\% and 95\% C.L. bounds on the EFT couplings are shown in Fig. \ref{b2}, which shows similar sensitivities of the EFT operators and modifiers as in Fig. \ref{b2}. The effects of EFT contributions to $Z$ decay branching and the assumption of custodial symmetry on these bounds are discussed in Appendix \ref{A} and Appendix \ref{E}, respectively. $Vh$ production has been studied in literature like \cite{Englert:2024ufr,Banerjee:2018bio} focusing on the boosted regime. However, both these analyses do not consider the light quark dipole operators which we include in our analysis. Considering the boosted regime is interesting as EFT effects are strongly evident there and hence the bounds on $c^{(3)}_{Hq}$ are slightly more stringent in both studies compared to our case. However, as we emphasized in our paper, the boosted regime may invalidate the EFT assumption, necessitating careful consideration. Studies like \cite{Banerjee:2019twi,Banerjee:2020vtm} focus on double differential observables with combination of multiple channels which results in further improvement in bound.
Similar to the $M_{llbb}$ distribution, other sensitive observables like $H_{T}$ (scalar sum of $p_{T}$ of visible particles) can also be used to perform this analysis. Another popular method is to train a ML based algorithm that separates the SM and SM+EFT contribution based on a set of discriminating observables. However, such model requires better segregation between the observables of the two classes which is evidently difficult even at HL-LHC once we abide by the EFT limit.

\subsection{Other future $pp$ colliders} \label{fpp}
\begin{figure}[h!]
	\centering
	\includegraphics[trim=1.5cm 0cm 0cm 1.5cm, clip, width = 0.525 \textwidth]{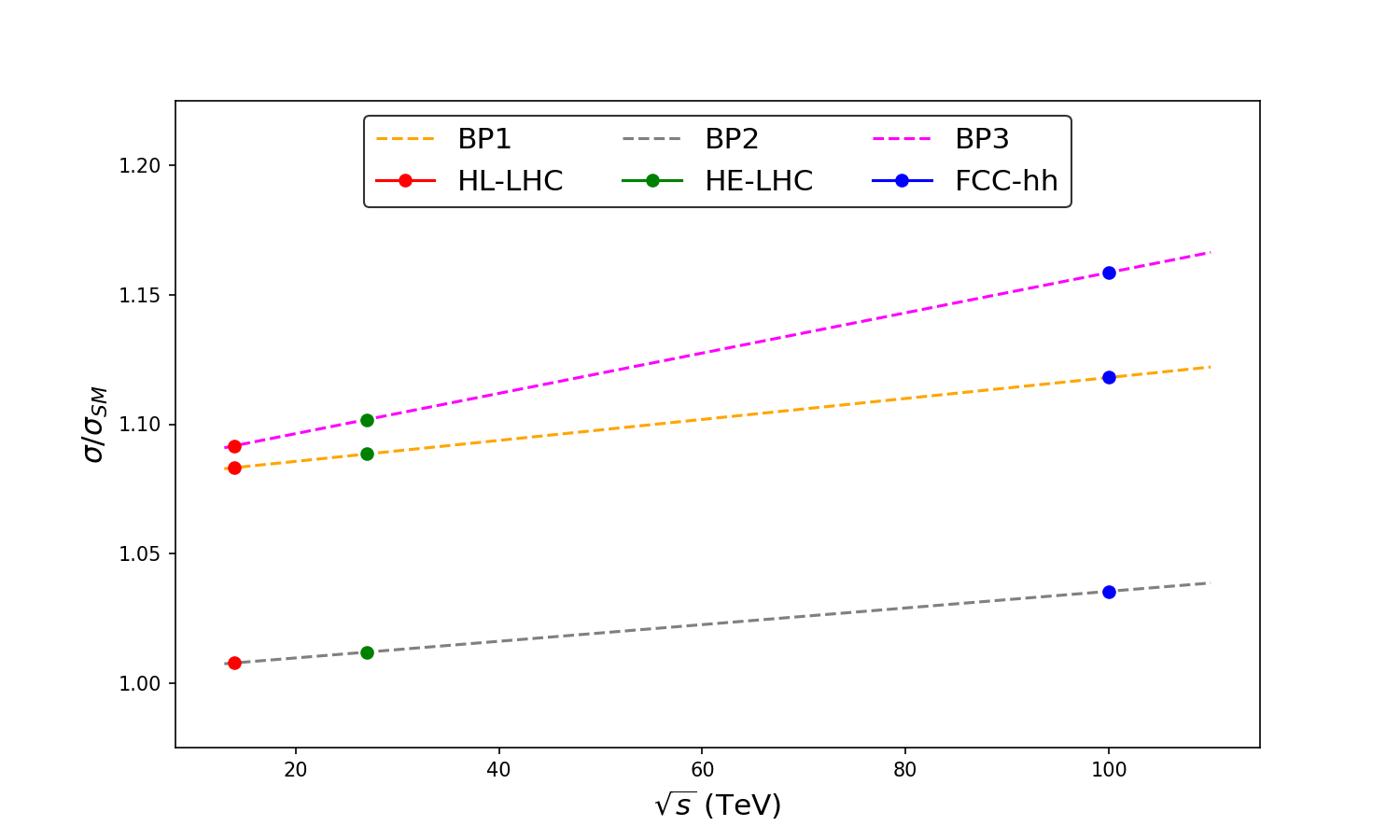} 
	\includegraphics[trim=0cm 0cm 1cm 1cm, clip, width= 0.425 \textwidth]{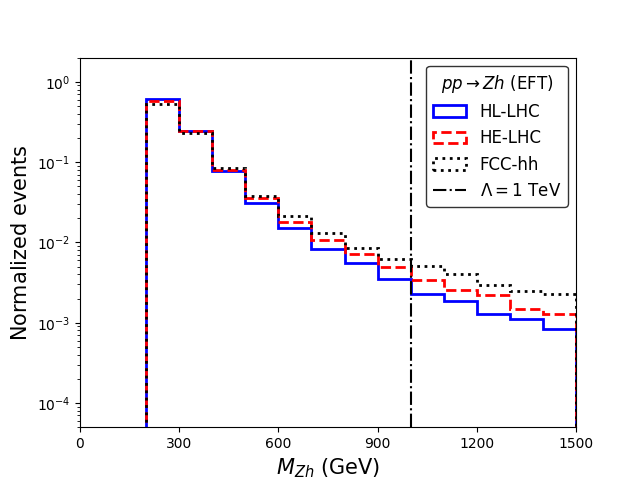}
	\caption{\textit{Left:} Variation in the ratio of $Zh$ production cross section to the SM cross section at different CoM energies for benchmarks $\{c_{uW}, c^{(3)}_{Hq}\}$: BP1 $\{0.05, 0.00\}$, BP2 $\{0.00, 0.05\}$ and BP3 $\{0.05, 0.05\}$. \textit{Right:} Invariant mass of $Zh$ for the parton level processes in presence of EFT operators at BP3 for different future $pp$ collider runs. \label{adx1}}
\end{figure}

As observed in Fig. \ref{fig:4} and \ref{fig:5}, the $Zh$ production cross section enhances for both the Higgs-current and dipole interactions, with increase in CoM energy. This motivates the possibility of better extraction of these NP contributions at higher CoM energy possible at future collider runs like the HE-LHC (expected to run at $\sqrt{s} = 27$ TeV), FCC-hh (expected to run at $\sqrt{s} = 100$ TeV), etc. $\frac{\sigma}{\sigma_{SM}}$ is plotted in Fig. \ref{adx1} (left) for different CoM energies ($\sqrt{s}$) at some of the benchmarks $\{c_{uW}, c^{(3)}_{Hq}\}$: BP1 $\{0.05, 0.00\}$, BP2 $\{0.00, 0.05\}$ and BP3 $\{0.05, 0.05\}$. All the benchmarks show enhancement in EFT effects, however we observe stronger enhancement for the Higgs-current interaction (BP2) in comparison to the dipole interaction (BP1). The combined presence of both the operators (BP3) shows the highest enhancement in cross-section. While these results show promise to better estimate the dipole operator in particular, the prospect of accelerating protons to such high energies remains a distant possibility.  Fig. \ref{adx1} (right) shows the validity of EFT limit, having the invariant mass peak well below $\Lambda=1$ TeV even in context of future $pp$ colliders with higher CoM energies. This happens due to the dominance of SM process, while larger Wilson coefficient would flatten the distribution having more events with higher invariant mass.

\section{Summary and conclusion} \label{s5}

We have presented an analysis of $Zh$ associated production at the LHC in presence of dimension six SMEFT operators and coupling modifiers. We choose a simple parametrisation to match the coupling modifiers that stem from dimension six SMEFT operators in Warsaw basis. We analysed the signal and background contributions at various benchmark points, which allows maximum cross-section within the allowed values of the EFT parameters and coupling modifiers adhering to the constraints appearing from existing collider and flavour constraints. We do both cut-based and BDT analysis using kinematic variables relevant for the process. NP scale is safely chosen at $\Lambda=1$ TeV, which abides by the effective limit having invariant mass distribution peak appearing below $\Lambda$.
  
A comparison of Table \ref{tab5} and \ref{tab5b} shows that the BDT analysis perform better than the cut based analysis presented in previous section as expected. One obtains almost factor 2 enhancement in the estimation of the significances for the various BP's considered in our analysis. This is due to the fact that BDT comes with better efficiency for the signal events even though the background efficiency remains almost the same. We have obtained more than $9\sigma$ ($5\sigma$) significance using BDT (cut based) analysis for all our chosen benchmark points. We further note that corresponding to our flavor scheme choice ($U(2)_{q} \times U(2)_{u} \times U(2)_{d}$) the corresponding background contributions have limited sensitivity to the EFT operators introduced here, hence we do not include them in our analysis. 

A $\chi^2$ analysis is performed in the two dimensional parameter space of the EFT operator coefficients as well as in the coupling modifiers. The bounds on the $c_{uW}-c^{(3)}_{Hq}$ plane is more constrained than those on $\kappa_{Z}-\kappa_{B}$ plane, owing to the fact that the differential distribution $M_{llbb}$ is extremely sensitive to EFT effects especially at the tail. Based on the plots in Figure \ref{fig:14}, it is evident that the EFT contributions significantly alters the bounds on the $\kappa_{Z}-\kappa_{B}$ plane. Similarly, coupling modifiers can also impact the bounds in the EFT coefficient planes as evident from Figure \ref{b2}. It is observed that even with the tight constraints on the operators, presence of Higgs-current operator $\mathcal{O}_{Hq}^{(3)}$ 
results in visible changes on the bounds in $\kappa_{Z}-\kappa_{B}$ plane. The individual impact of the dipole operator $\mathcal{O}_{uW}$ is relatively small and do not show any significant deviation from the SM predictions. This is mainly because, they do not interfere with the SM process and the contributions are obtained only at $1/\Lambda^4$ order. However, as discussed in section \ref{fpp}, significant increase in sensitivity is expected with increase in the CoM energy, possible at future high energy collider runs. However, the contributions from $\mathcal{O}_{Hq}^{(3)}$ will still dominate. 

In this analysis, we have omitted CP odd contributions to the dipole operators, primarily because they are strongly bounded from EDM measurements. Therefore, it is difficult to detect their effects via signal cross-section. Dedicated CP violating observables may decipher a disentanglement from that of a CP even case. This lies outside the scope of the present work, but will be taken up in future.


\acknowledgments

Subhaditya would like to acknowledge the grant CRG/2019/004078 from DST-SERB, Govt. of India.

\appendix

\section{Constraints on operators} \label{D}
The constraints on dipole (other than $c_{uW}$) and Higgs-current (other than $c^{(3)}_{Hq}$) operator coefficients from existing studies are listed in Table \ref{tab:031} and \ref{tab:03a1}, respectively.

\begin{table}[htbp]
	\centering
	\begin{tabular}{|c|c|c|c|}
		\hline
		Coefficient & Bound & C.L. & Flavor Scheme [Source] \\ \hline
		\multirow{6}{*}{$c_{uB}$} & [-0.602, 0.602] & 95\% & $U(3)_{q} \times U(3)_{u} \times U(3)_{d}$ \cite{Boughezal:2021tih} \\ 
		&  [-0.780, 0.780] & 95\% & $U(2)_{q} \times U(2)_{u} \times U(2)_{d}$ \cite{daSilvaAlmeida:2019cbr} \\
		& [-0.199, 0.093] & 95\% & $U(1)_{Q} \times U(1)_{t}$ \cite{Ethier:2021bye} \\
		& [-0.036, 0.324] & 90\% & $U(3)_{q} \times U(3)_{u} \times U(3)_{d}$ \cite{Grunwald:2023nli} \\
		& [-0.430, 0.284] & 95\% & $U(1)_{Q} \times U(1)_{t}$ \cite{Buckley:2015lku} \\
		& [-4.500, 1.200] & 95\% & $U(1)_{Q} \times U(1)_{t}$ \cite{Ellis:2020unq} \\
		\hline
		\multirow{2}{*}{$c_{dW}$} & [-0.484, 0.484] & 95\% & $U(3)_{q} \times U(3)_{u} \times U(3)_{d}$ \cite{Boughezal:2021tih} \\
		& [-0.360, 0.360] & 95\% & $U(2)_{q} \times U(2)_{u} \times U(2)_{d}$ \cite{daSilvaAlmeida:2019cbr} \\
		\hline
		\multirow{2}{*}{$c_{dB}$} & [-0.769, 0.769] & 95\% & $U(3)_{q} \times U(3)_{u} \times U(3)_{d}$ \cite{Boughezal:2021tih} \\
		& [-0.960, 0.960] & 95\% & $U(2)_{q} \times U(2)_{u} \times U(2)_{d}$ \cite{daSilvaAlmeida:2019cbr} \\ 
		\hline
	\end{tabular}
	\caption{Constraints on dipole operator coefficients from existing studies, see the references for details. \label{tab:031}}
\end{table}

\begin{table}[htbp]
	\centering
	\begin{tabular}{|c|c|c|c|}
		\hline
		Coefficient & Bound  & C.L. & Flavor Scheme [Source] \\ \hline
		\multirow{8}{*}{$c^{(1)}_{Hq}$} & [-2.659, 0.381] & 95\% & $U(2)_{q} \times U(2)_{u} \times U(3)_{d}$ \cite{Ethier:2021bye} \\
		& [-1.147, 1.585] & 95\% & $U(1)_{Q} \times U(1)_{t}$ \cite{Ethier:2021bye} \\
		& [-0.260, 0.115] & 68\% & $U(3)_{q} \times U(3)_{u} \times U(3)_{d}$ \cite{ThomasArun:2023wbd} \\
		& [-1.600, 0.430] & 95\% & $U(3)_{q} \times U(3)_{u} \times U(3)_{d}$ \cite{ATLAS:2021kog} \\
		& [-0.048, 0.021] & 90\% & $U(3)_{q} \times U(3)_{u} \times U(3)_{d}$ \cite{Grunwald:2023nli} \\
		& [-0.100, 0.140] & 95\% & $U(2)_{q} \times U(2)_{u} \times U(3)_{d}$ \cite{Ellis:2020unq} \\
		& [-0.031, 0.049] & 95\% & $U(1)_{Q} \times U(1)_{t}$ \cite{Ellis:2020unq} \\
		& [-0.023, 0.047] & 95\% & $U(2)_{q} \times U(2)_{u} \times U(3)_{d}$ \cite{Anisha:2021hgc} \\
		\hline
		\multirow{7}{*}{$c_{Hu}$} & [-0.458, 0.375] & 95\% & $U(2)_{q} \times U(2)_{u} \times U(3)_{d}$ \cite{Ethier:2021bye} \\
		& [-1.038, 0.449] & 68\% & $U(3)_{q} \times U(3)_{u} \times U(3)_{d}$ \cite{ThomasArun:2023wbd} \\
		& [-0.060, 0.036] & 90\% &  $U(3)_{q} \times U(3)_{u} \times U(3)_{d}$ \cite{Grunwald:2023nli} \\
		& [-0.190, 0.100] & 68\% & $U(3)_{q} \times U(3)_{u} \times U(3)_{d}$ \cite{ATLAS:2020fcp} \\
		& [-0.075, 0.073] & 95\% & $U(2)_{q} \times U(2)_{u} \times U(3)_{d}$ \cite{Ellis:2020unq} \\
		& [-1.200, 2.900] & 95\% & $U(1)_{Q} \times U(1)_{t}$ \cite{Ellis:2020unq} \\
		& [-0.056, 0.081] & 95\% & $U(2)_{q} \times U(2)_{u} \times U(3)_{d}$ \cite{Anisha:2021hgc} \\
		\hline
		\multirow{5}{*}{$c_{Hd}$} & [-0.187, 0.229] & 95\% & $U(2)_{q} \times U(2)_{u} \times U(3)_{d}$ \cite{Ethier:2021bye} \\		
		& [-0.520, 0.267] & 68\% & $U(3)_{q} \times U(3)_{u} \times U(3)_{d}$ \cite{ThomasArun:2023wbd} \\
		& [-2.600, 8.300] & 95\% & $U(3)_{q} \times U(3)_{u} \times U(3)_{d}$ \cite{ATLAS:2021kog} \\
		& [-0.130, 0.071] & 95\% & $U(2)_{q} \times U(2)_{u} \times U(3)_{d}$ \cite{Ellis:2020unq} \\
		& [-0.150, 0.040] & 95\% & $U(2)_{q} \times U(2)_{u} \times U(3)_{d}$ \cite{Anisha:2021hgc} \\
		\hline
	\end{tabular}
	\caption{Constraints on Higgs-current operator coefficients from existing studies, see the references for details. \label{tab:03a1}}
\end{table}

Bounds on the imaginary part of dipole operators from neutron EDM \cite{Kley:2021yhn} (assuming $U(2)_{q} \times U(2)_{u} \times U(2)_{d}$ flavor scheme) are shown below:
\begin{equation}
	\begin{split}
	\mathfrak{Im}(c_{uB}) & < 0.000368 \\
	\mathfrak{Im}(c_{dW}) & < 0.000002 \\
	\mathfrak{Im}(c_{dB}) & < 0.000001 \\
	\end{split}
\end{equation}

\section{Details of BDT analysis} \label{C}

\begin{figure}[h!]
	\centering
	\includegraphics[width = 0.475 \textwidth]{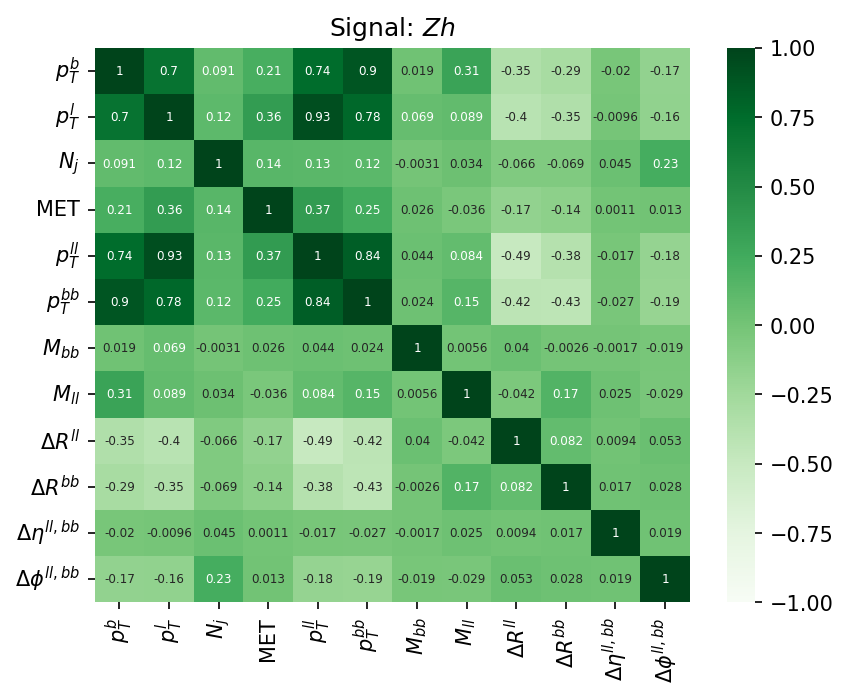}
	\includegraphics[width = 0.475 \textwidth]{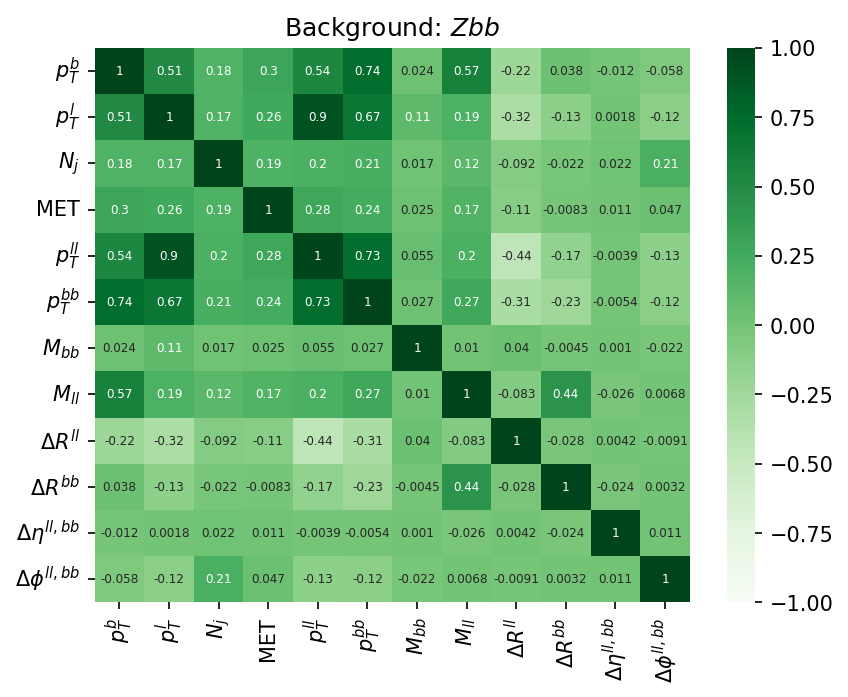}
	\includegraphics[width = 0.475 \textwidth]{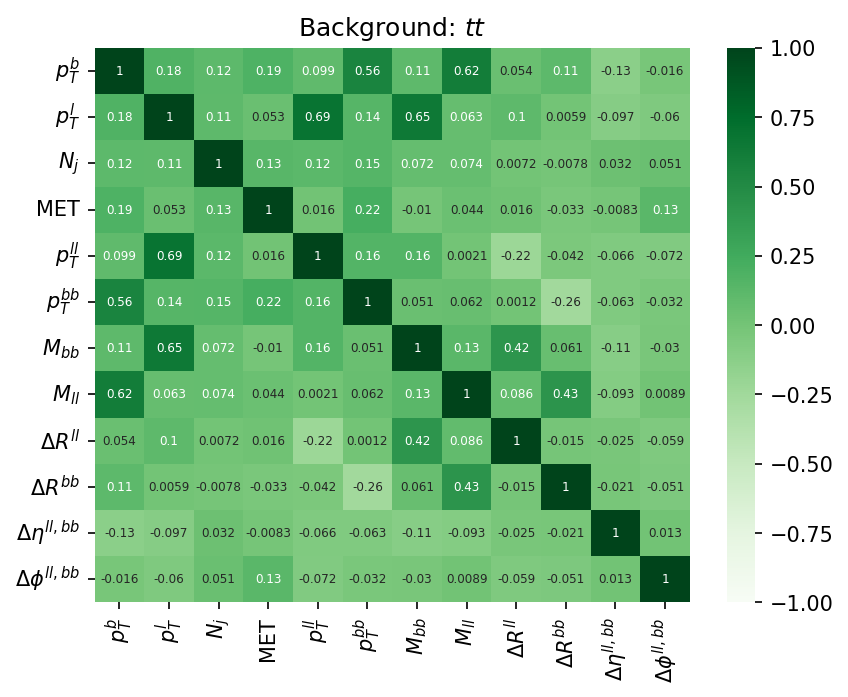}
	\includegraphics[width = 0.475 \textwidth]{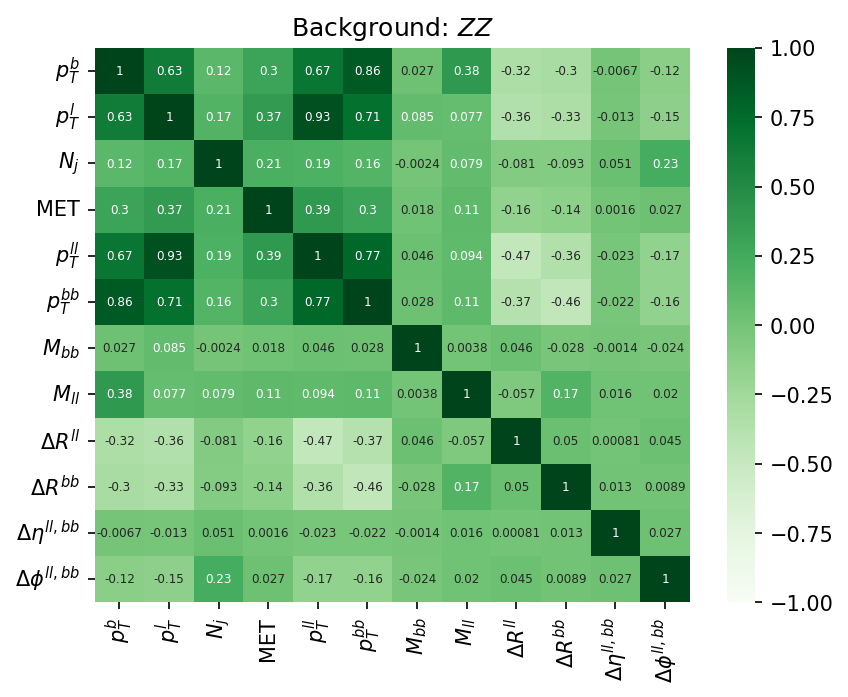}
	\caption{\textit{Top Left:} Correlation matrix for the features used in training the XGBoost model for $Zh$ process. \textit{Top Right:} $Zbb$ process. \textit{Bottom Left:} $t\overline{t}$ process. \textit{Bottom Right:} $ZZ$ process. \label{b3}}
\end{figure}

Feature selection is a integral and significant part of any machine learning analysis. Presence of too many redundant features causes a model to learn unnecessary characteristics, as such we expect the features to have little dependence on one another. The correlation matrix captures the interdependnce among the features very accurately. For our BDT model, the feature correlation heatmap is shown in Figure \ref{b3}. Darker patches refer to high correlation. Apart from features likes $p_{T}^{l}$ and $p_{T}^{ll}$, $p_{T}^{b}$ and $p_{T}^{bb}$, where correlation is expected, most of the features show little to no correlation, enhancing the model's learnability against general data. Apart from feature selection, another important aspect of model tuning is the hyperparameter optimization. Proper tuning prevents the model from overfitting, resulting in better performance on new data. The optimized hyperparameters for the model are detailed in Table \ref{hp}.

\begin{table}[h!]
	\centering
	\begin{tabular}{|l|c|}
		\hline
		Hyperparameters for the XGBClassifier model & Optimal value \\ \hline
		Number of boosting rounds (\texttt{n\_estimators}) & 50 \\
		Maximum depth of a tree (\texttt{max\_depth}) & 5 \\
		Learning rate for the model (\texttt{learning\_rate}) & 0.15 \\
		Fraction of samples used each tree (\texttt{subsample}) & 0.8 \\
		Fraction of features used for each tree (\texttt{colsmaple\_bytree}) & 1.0 \\
		Minimum sum of instance weight in a child (\texttt{min\_child\_weight}) & 50 \\
		Minimum loss reduction for further partition (\texttt{gamma}) & 0.75 \\
		L1 regularization term on weights (\texttt{reg\_alpha}) & 8 \\
		L2 regularization term on weights (\texttt{reg\_lambda}) & 60 \\
		Maximum step size for each iteration (\texttt{max\_delta\_step}) & 2 \\ \hline
	\end{tabular}
	\caption{The optimal hyperparameter values for the model mined using \texttt{GridSearchCV}. The \texttt{objective} hyperparameter is fixed at \texttt{"binary:logistic"}. \label{hp}}
\end{table}

\section{Differential distributions} \label{B}
\begin{figure}[htbp]
	\centering
	\includegraphics[trim=0 0 1cm 1cm, clip, width=0.4 \textwidth]{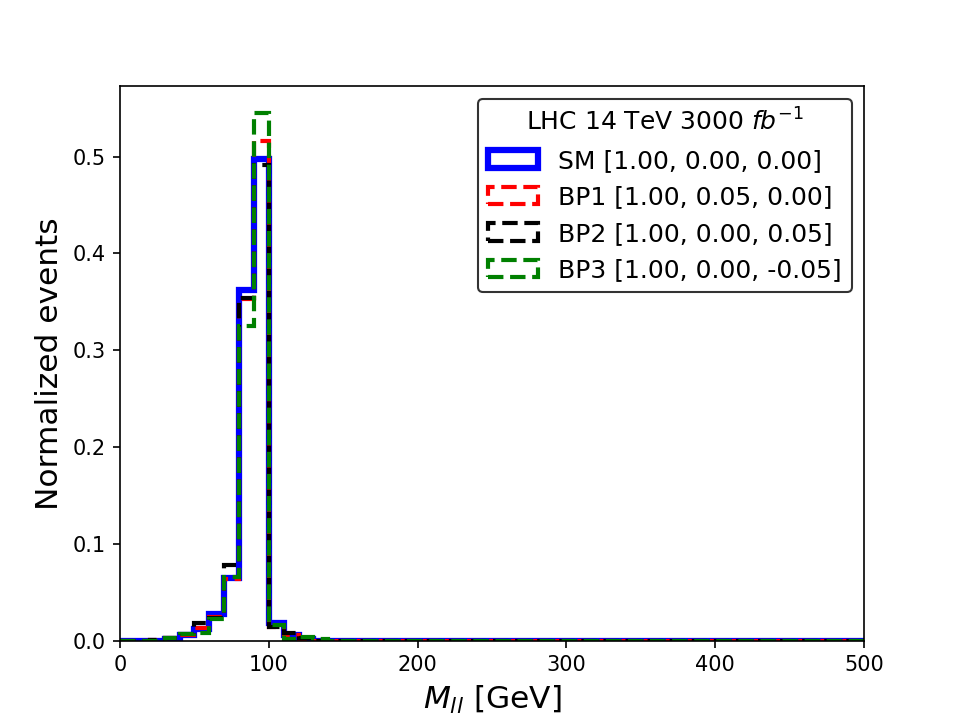}
	\includegraphics[trim=0 0 1cm 1cm, clip, width=0.4 \textwidth]{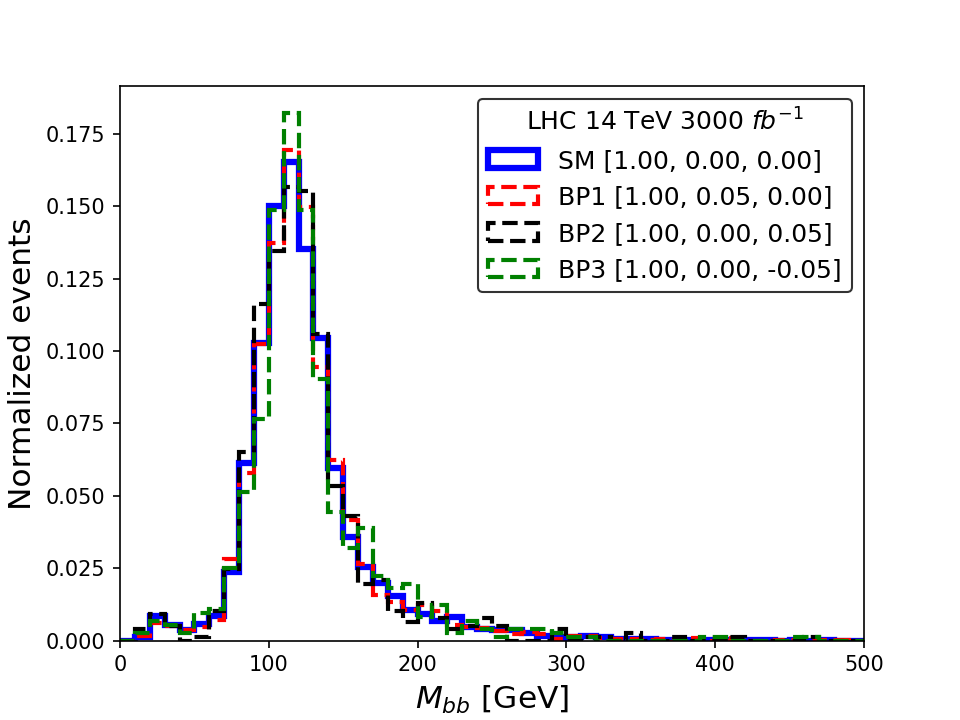}
	\includegraphics[trim=0 0 1cm 1cm, clip, width=0.4 \textwidth]{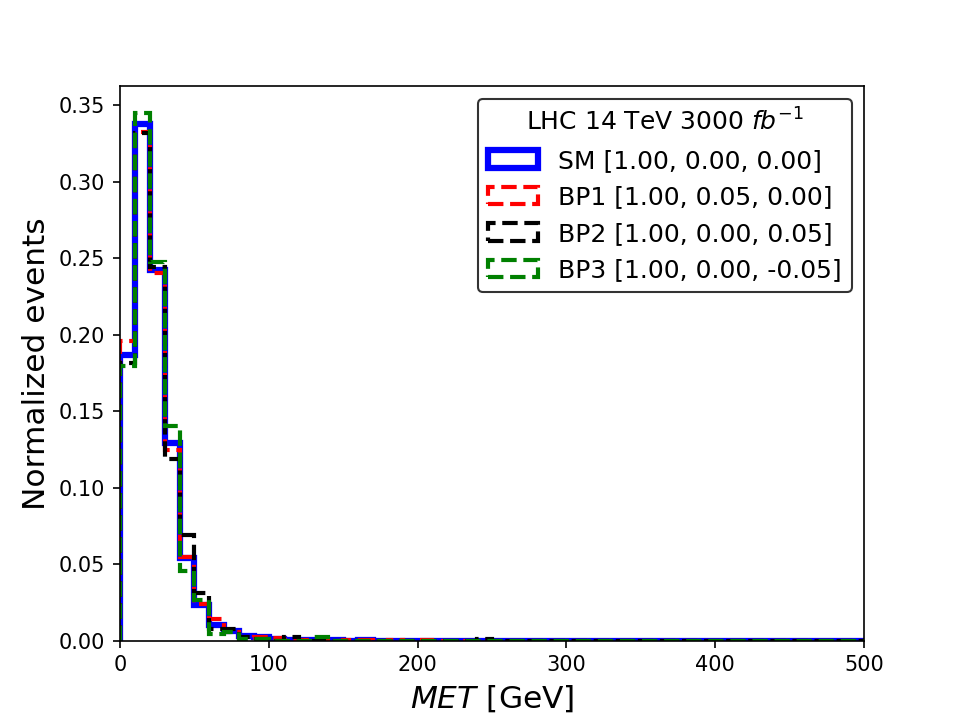}
	\caption{\textit{Top Left:} Invariant mass of dilepton, \textit{Top Right:} invariant mass of di-bjet and \textit{Bottom:} missing transverse energy corresponding to $Z(l^{+}l^{-})h(b \Bar{b})$ at $\sqrt{s}=$14 TeV at different EFT benchmark 
	points shown in Table \ref{tab5}.\label{fig:9a}}
\end{figure}

\begin{figure}[h!]
	\centering
	\includegraphics[trim=0 0 1cm 1cm, clip, width = 0.4 \textwidth]{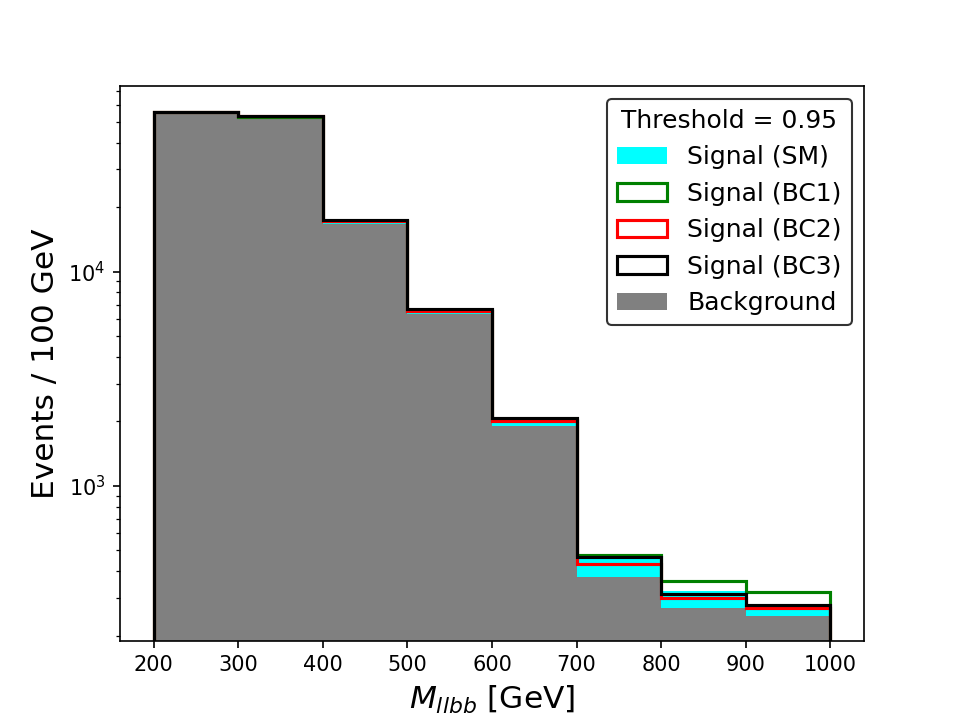}
	\includegraphics[trim=0 0 1cm 1cm, clip, width = 0.4 \textwidth]{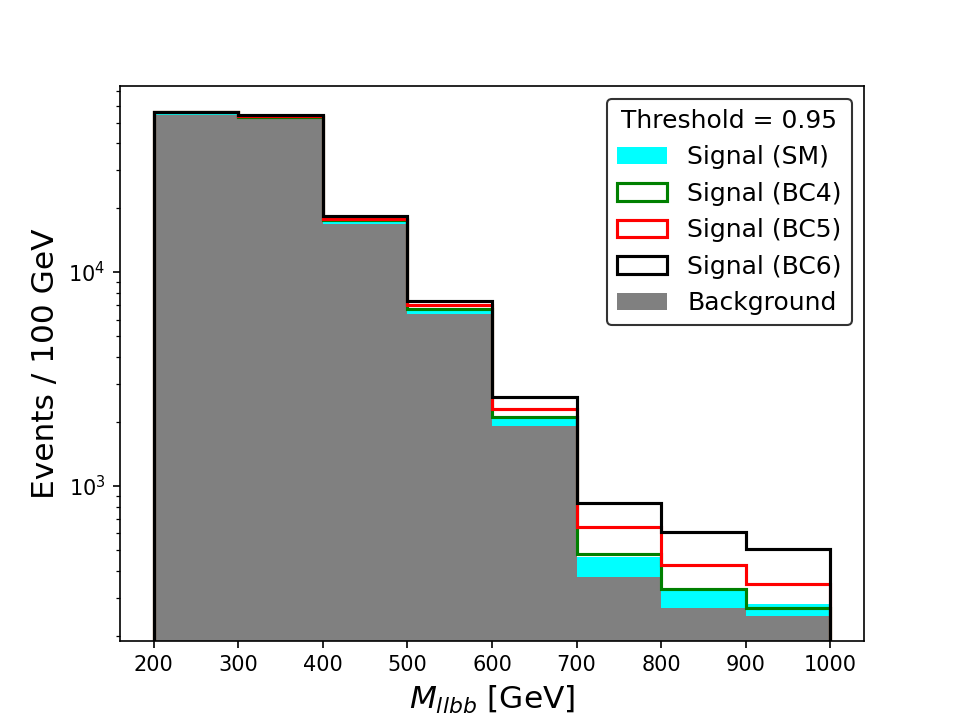}
	\includegraphics[trim=0 0 1cm 1cm, clip, width = 0.4 \textwidth]{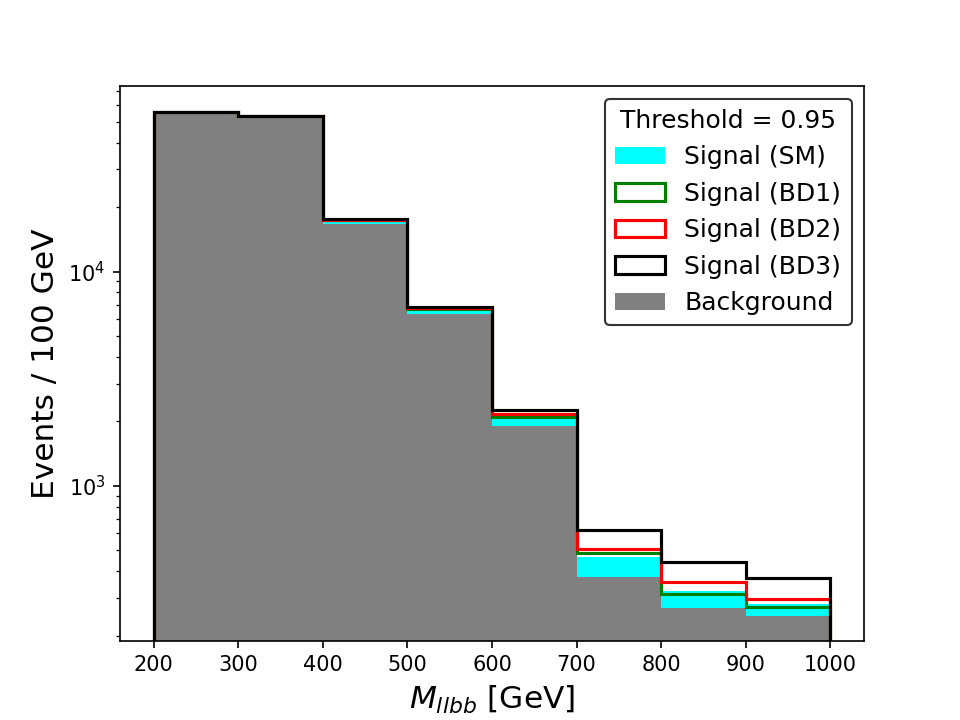}
	\caption{$M_{llbb}$ distributions for different EFT benchmarks post threshold choice of 0.95. \label{fig:dd}}
\end{figure}

Kinematic distributions for different EFT benchmark points are shown in Figure \ref{fig:9a}.
The $M_{llbb}$ distribution post BDT score threshold choice of 0.95 are shown in Figure \ref{fig:dd}. The benchmarks $\{c_{uW}, c^{(3)}_{Hq}\}$ are: \textit{Left:} BC1 $\{0.00, -0.20\}$, BC2 $\{0.00, -0.10\}$ and BC3 $\{0.00, -0.01\}$; \textit{Center:} BC4 $\{0.00, 0.01 \}$, BC5 $\{0.00, 0.10 \}$ and BC6 $\{0.00, 0.20 \}$; \textit{Right:} BD1 $\{0.01, 0.00 \}$, BD2 $\{0.10, 0.00 \}$ and BD3 $\{0.20, 0.00 \}$.

\section{$Z$ branching correction} \label{A}
The correction to Z decay width in presence of $\mathcal{O}_{uW}$ and $\mathcal{O}^{(3)}_{Hq}$ can be parametrized as:
\begin{equation}
	\Gamma_{Z} = \Gamma^{SM}_{Z} \left[1 + 0.0047 \left(c_{uW}\right)^{2} + 0.0099 \left(c^{(3)}_{Hq}\right)^{2} + 0.0976 \left(c^{(3)}_{Hq}\right) \right]
\end{equation}
The correction to $Z \rightarrow l^{+} l^{-}$ branching is
$\mathcal{O}^{(3)}_{Hq}$ can be parametrized as:
\begin{equation}
	\left(B.R.\right)_{Zll} = \frac{\Gamma_{Zll}}{\Gamma_{Z}} = \frac{\left(B.R.\right)^{SM}_{Zll}}{\left[1 + 0.0047 \left(c_{uW}\right)^{2} + 0.0099 \left(c^{(3)}_{Hq}\right)^{2} + 0.0976 \left(c^{(3)}_{Hq}\right) \right]}
\end{equation}

\begin{figure}[htbp]
	\centering
	\includegraphics[trim=0 0 0 0, clip, width = 0.4 \textwidth]{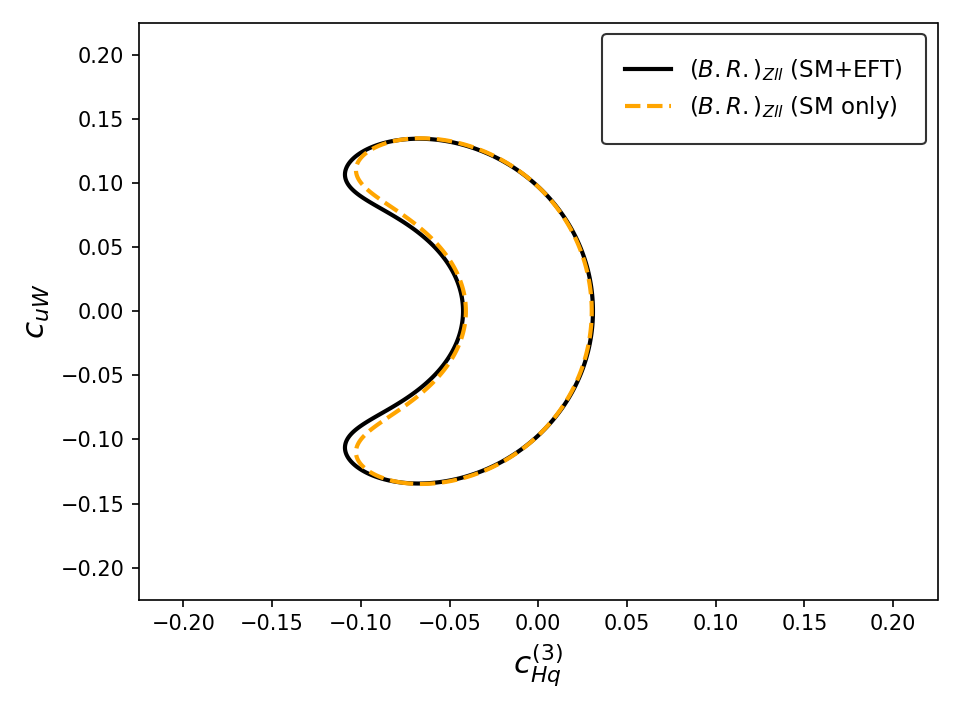}
	\includegraphics[trim=0 0 0 0, clip, width = 0.4 \textwidth]{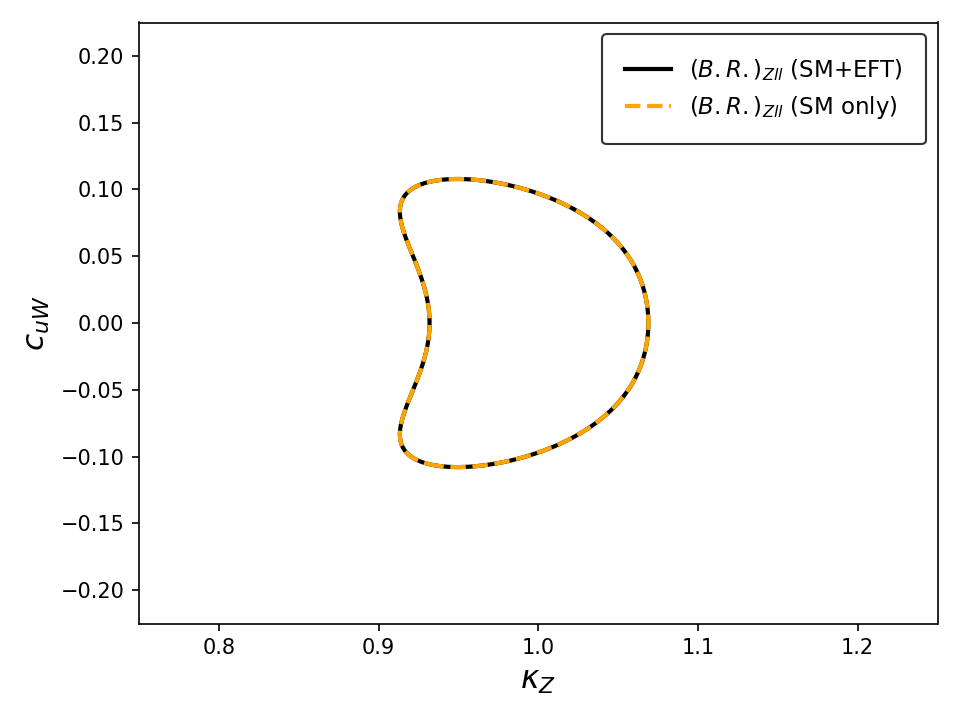}
	\includegraphics[trim=0 0 0 0, clip, width = 0.4 \textwidth]{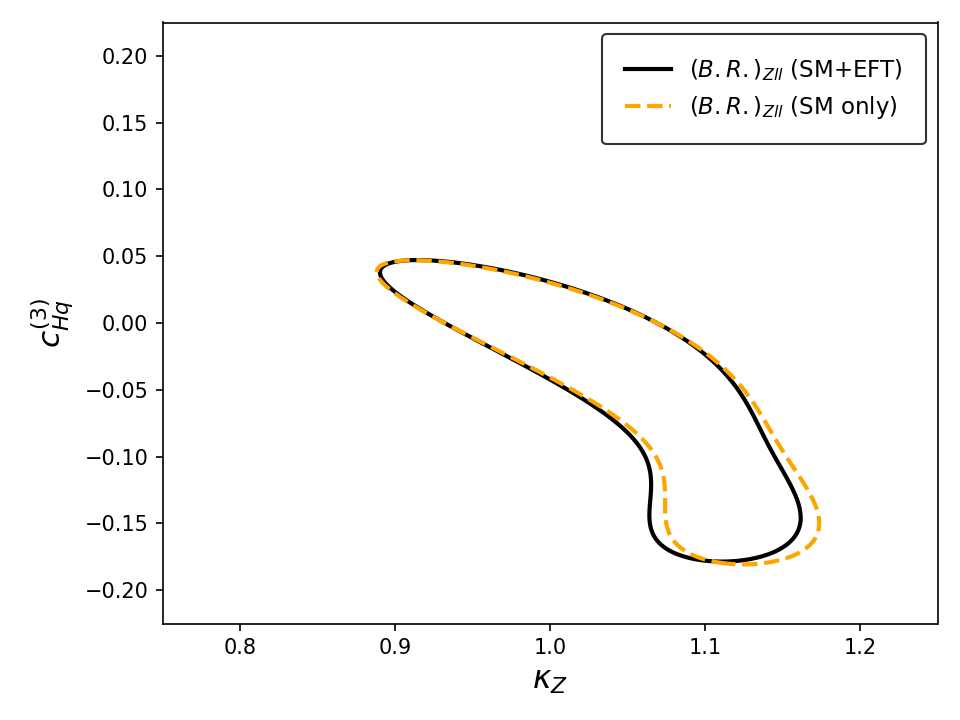}
	\caption{Shift in 68\% C.L. limits depending on whether EFT corrections to the Z branching are included or not. \label{zd}}
\end{figure}

Figure \ref{zd} shows how the 68\% C.L. limits change depending on whether EFT corrections to the Z branching are included or not. The effect is not very strong, but it is taken into account in our analysis.

\section{The effect of custodial symmetry} \label{E}
\begin{figure}[htbp]
		\centering
		\includegraphics[trim=0 0 0 0, clip, width = 0.4\textwidth]{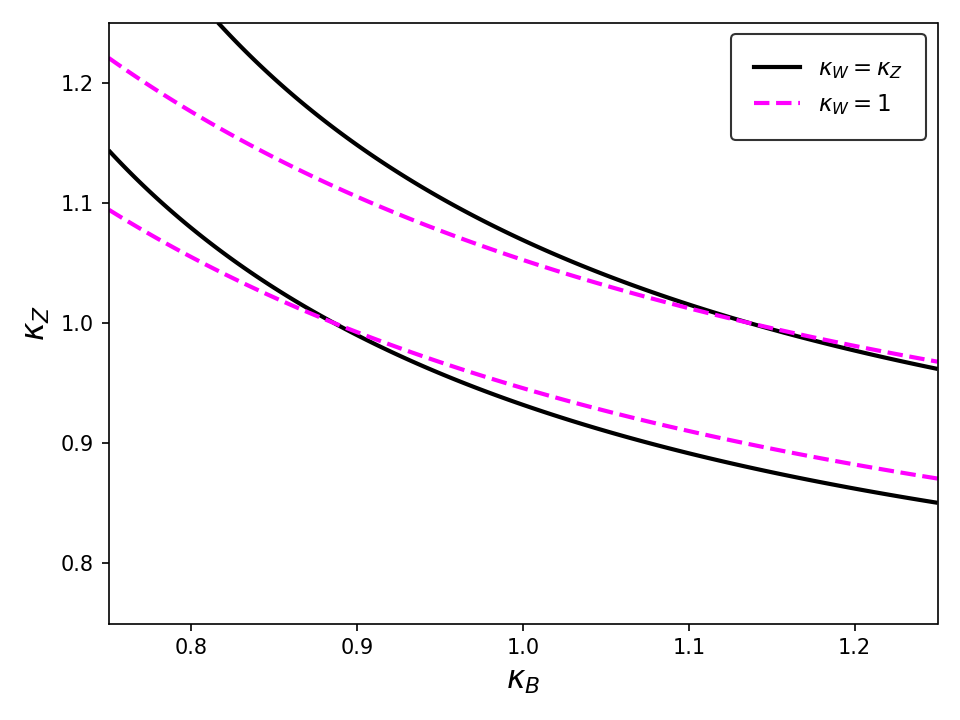}
		\includegraphics[trim=0 0 0 0, clip, width = 0.4\textwidth]{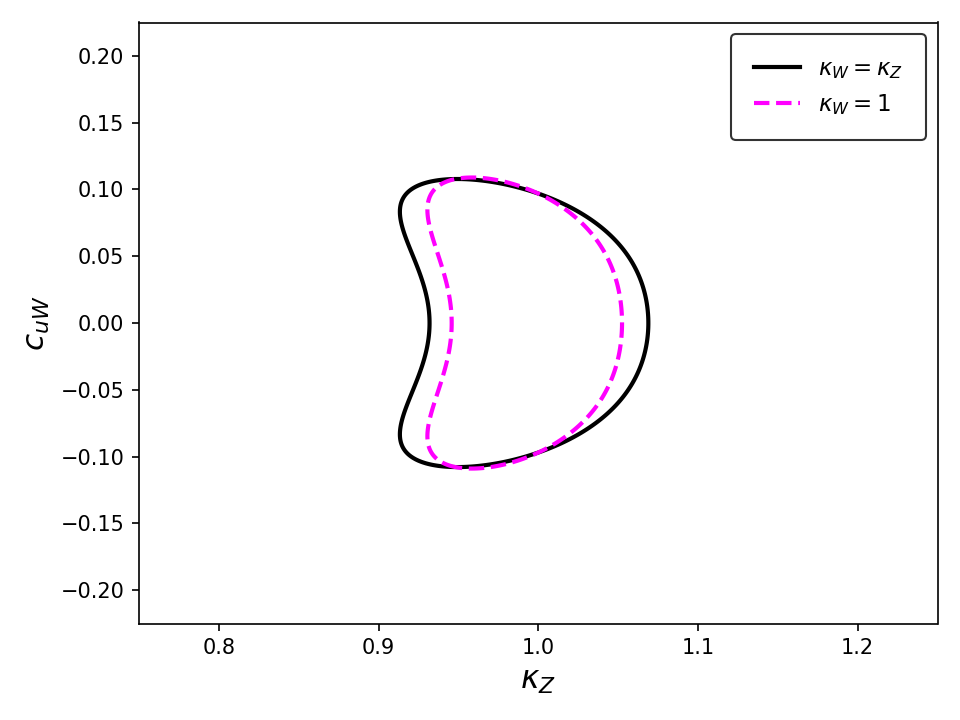}
		\includegraphics[trim=0 0 0 0, clip, width = 0.4\textwidth]{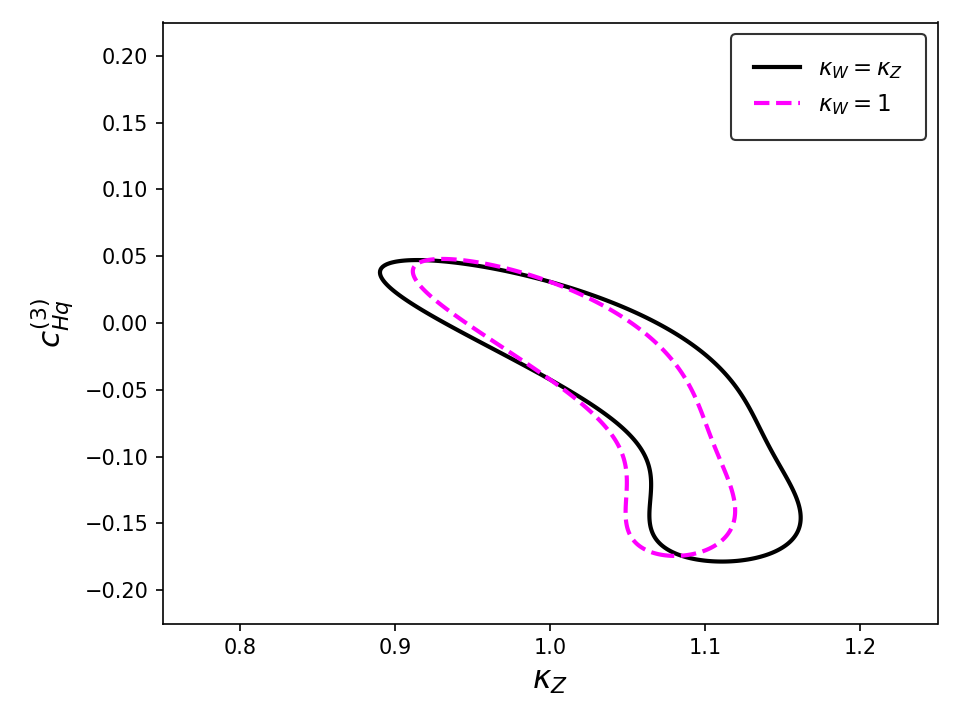}
		\caption{Shift in 68\% C.L. limits depending on whether custodial symmetry is assumed. \label{kw}}
	\end{figure}

In Section \ref{s3}, we imposed the custodial symmetry condition $\kappa_W = \kappa_Z$. However, this assumption can be relaxed. In this section, we explore how the 68\% C.L. limits change under two scenarios: (i) with custodial symmetry, where $\kappa_W = \kappa_Z$ (as discussed in Section \ref{s3}), and (ii) without custodial symmetry, where $\kappa_W \neq \kappa_Z$ and $\kappa_W = 1$, assuming an SM-like coupling of the Higgs to the $W$ boson while treating $\kappa_Z$ as a free parameter. For the second scenario, the Higgs production signal strength via the $Zh$ channel is given by:
	\begin{equation} \label{eex4}
		\mu^{Zh}_{hbb} = \frac{(\sigma^{Zh} \times (B.R.)_{hbb})}{(\sigma^{Zh} \times (B.R.)_{hbb})_{SM}} = \frac{\kappa_Z^2 \kappa_B^2}{0.3929 + 0.5809 \kappa_B^2 + 0.0262 \kappa_Z^2},.
	\end{equation}
Figure \ref{kw} illustrates how the 68\% C.L. limits are affected by the assumption of custodial symmetry. The assumption significantly impacts our bounds, particularly when $\kappa_Z$ deviates from 1.
	
\newpage
\bibliographystyle{JHEP}
\bibliography{biblio.bib}

\end{document}